\pdfoutput=1
\documentclass[10pt,twocolumn,aps,prl,superscriptaddress,floatfix,notitlepage]{revtex4-1}

\usepackage[utf8]{inputenc}
\usepackage[T1]{fontenc}
\usepackage[british]{babel}
\usepackage{lmodern}
\usepackage{graphicx}
\usepackage{amsmath}
\usepackage{amssymb}
\usepackage{bm}
\usepackage[dvipsnames]{xcolor}
\usepackage{csquotes}
\usepackage{tabularx}
\usepackage{mathtools}
\usepackage{microtype}
\usepackage[load=physical,load=abbr]{siunitx}
\usepackage[version=4]{mhchem}
\usepackage{bbm}
\usepackage{booktabs,siunitx}

\usepackage{braket}
\usepackage{mathtools}

\usepackage[normalem]{ulem}

\usepackage[colorlinks, citecolor={blue!50!black}, urlcolor={blue!50!black}, linktoc=all, pdfencoding=auto]{hyperref}
\usepackage{bookmark}

\def\ii{\ensuremath{\mathrm{i}}}

\def\ZZ{\ensuremath{\mathbb{Z}}}

\def\br{\ensuremath{\bm{r}}}
\def\bc{\ensuremath{\bm{c}}}

\def\Tr{\ensuremath{\operatorname{Tr}}}

\graphicspath{{figures/}}

\newcommand{\alloy}{\ce{Pb_{1-x}Sn_{x}Te}}

\renewcommand{\paragraph}[1]{%
  \par\refstepcounter{paragraph}
  \paragraphmark{#1}
  \addcontentsline{toc}{paragraph}{\protect\numberline{\theparagraph}#1}
}
\newcommand{\co}[1]{\paragraph{#1}}

\newcommand{\startsection}[1]{%
  \par\refstepcounter{section}
  \sectionmark{#1}
  \addcontentsline{toc}{section}{\protect\numberline{\thesection}#1}
  \emph{#1. \textemdash}%
}

\begin{document}

\author{D{\'a}niel Varjas}
\email[Electronic address: ]{dvarjas@gmail.com}
\affiliation{QuTech, Delft University of Technology, P.O. Box 4056, 2600 GA Delft, The Netherlands}
\affiliation{Kavli Institute of Nanoscience, Delft University of Technology, P.O. Box 4056, 2600 GA Delft, The Netherlands}

\author{Michel Fruchart}
\affiliation{Instituut-Lorentz, Universiteit Leiden, P.O. Box 9506, 2300 RA Leiden, The Netherlands}
\affiliation{James Franck Institute and Department of Physics, University of Chicago, Chicago IL 60637, USA}

\author{Anton R. Akhmerov}
\affiliation{Kavli Institute of Nanoscience, Delft University of Technology, P.O. Box 4056, 2600 GA Delft, The Netherlands}

\author{Pablo M. Perez-Piskunow}
\email[Electronic address: ]{pablo.perez.piskunow@gmail.com}
\affiliation{Kavli Institute of Nanoscience, Delft University of Technology, P.O. Box 4056, 2600 GA Delft, The Netherlands}
\affiliation{Catalan Institute of Nanoscience and Nanotechnology (ICN2), CSIC and BIST, Campus UAB, Bellaterra, 08193 Barcelona, Spain}

\title{Computation of topological phase diagram of disordered \texorpdfstring{$\text{Pb}_{1-x}\text{Sn}_{x}\text{Te}$}{PbSnTe}\texorpdfstring{\\}{}using the kernel polynomial method}

\begin{abstract}
We present an algorithm to determine topological invariants of inhomogeneous systems, such as alloys, disordered crystals, or amorphous systems.
Based on the kernel polynomial method, our algorithm allows us to study samples with more than $10^7$ degrees of freedom.
Our method enables the study of large complex compounds, where disorder is inherent to the system.
We use it to analyse \alloy{} and tighten the critical concentration for the phase transition.
Moreover, we obtain the topological phase diagram for related alloys in the family of three-dimensional mirror Chern insulators.

\end{abstract}
\maketitle

\co{People search for topological materials using high throughput algorithms.}
\startsection{Introduction}
Topological materials have attracted continuing interest from both the fundamental physics and the material science communities for the last decade~\cite{qi2011topological,Chiu2016}.
The program to theoretically classify non-interacting crystalline insulators has been completed, tabulating possible topological phases in all space groups~\cite{Kruthoff2017, po2017classification, Bradlyn2017}.
Recent efforts focus on automated high-throughput methods to discover and classify new topological materials~\cite{Yang2012,Curtarolo2013,Zhang2018b}, culminating in the production of comprehensive databases of topological insulators and semimetals~\cite{Tang2018a,Tang2018b,Zhang2018,Vergniory2018,Zhang2018b}.

\co{Alloys are important.}
However, not all topological insulators are compounds with perfect stoichiometry.
The first three-dimensional topological insulator to be predicted~\cite{Fu2007} and experimentally realized~\cite{Hsieh2008} was an alloy---\ce{Bi_{x}Sb_{1-x}}.
These systems are usually studied using the virtual crystal approximation or the coherent potential approximation, which approximate an alloy by a perfect crystal~\cite{Chadov2013,Sante2015}.
This approach ignores the intrinsic disorder in alloys, and it is insufficient to explain topological transitions that appear at strong disorder~\cite{Li2009, Groth2009}, or accurately find critical concentrations.

\co{Alloys need large supercell, this is expensive.}
The topological invariant converges to its bulk value in samples larger than the localization length $\xi$.
This is the main limitation in resolving topological phase transitions as $\xi$ diverges.
Therefore, the asymptotic scaling of the computational cost with $\xi$ is the main distinction between different numerical approaches.

\co{Several alternatives exist, but their performance is unsuitable for 3D.}
Available methods to compute topological invariants either apply the momentum-space Berry curvature formalism to periodic systems with a disordered supercell, or use a real space formulation on a large finite sample~\cite{Kitaev2006,Essin2007,Loring2010,Bianco2011,Loring2015,Prodan2013,Prodan2016,Akagi2017,Katsura2018}.
However, these methods involve solving at least one eigenvalue equation with size equal to the number of degrees of freedom, resulting in the complexity $\xi^{3d}$ in $d$ dimensions.
This restricts the applicability of such methods to small system sizes, especially in three dimensions (3D).
To our knowledge, the most efficient method in 3D is the scattering matrix approach~\cite{Fulga2014}, with a complexity scaling as $\xi^{3(d-1)}$, allowing for maximum sizes of $5\times10^5$ degrees of freedom.

\co{We develop a new method based on KPM.}
We present an algorithm to efficiently identify topological phases of strongly disordered systems using the kernel polynomial method (KPM)~\cite{Weisse2006,Rappoport2015,Carvalho2018}, an approximation based on a polynomial expansion of the quantities of interest.
All topological properties of a non-interacting system of electrons are encoded in the projector on the occupied states (spectral projector), which is efficiently approximated using KPM.
Our algorithm builds on the method of topological markers~\cite{Bianco2011} to construct a topological invariant as the trace of an operator.
Since topological invariants are integers, it is sufficient to reduce the statistical uncertainty below $1/2$ to obtain the exact value.
In particular, the stochastic evaluation of traces~\cite{Weisse2006} from a small number of random vectors, combined with KPM, is suited for this task.
With $\xi^{d+1}$ scaling of the computational effort, it is the most efficient method in three dimensions.

\begin{figure}[ht]
    \includegraphics[width=0.95\columnwidth]{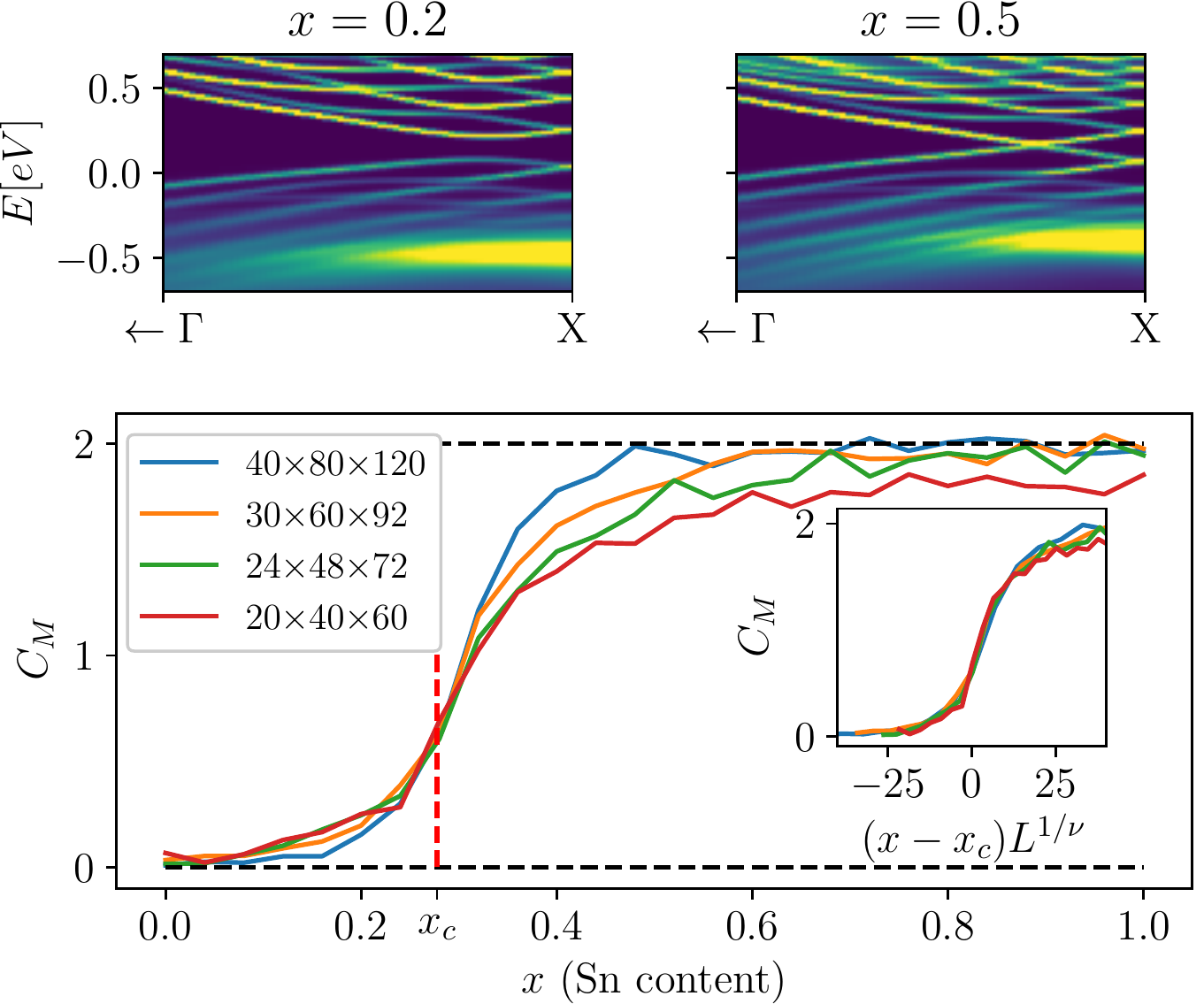}
    \caption{Top: surface spectra of a $20\times 80\times 80$ sample of \alloy{} in the trivial (left) and topological (right) phase.
    The presence of a gapless surface Dirac cone indicates the mirror Chern insulator phase.
    Bottom: transition between trivial and mirror Chern phase when varying $x$ for \alloy{} calculated using our method with various system sizes.
    Inset: Finite size collapse of the curves with $x_c \simeq \num{0.28(3)}$ and $\nu \simeq \num{0.9(6)}$.}
    \label{fig:surface_and_mirror_chern}
\end{figure}

\co{We use this method to analyze large samples of real materials.}
As a concrete example, we apply our method to lead tin telluride \alloy{} alloys---three-dimensional topological crystalline insulators characterized by a mirror Chern number.
Thanks to the efficiency of our algorithm, we analyze 3D systems with linear sizes of over a hundred lattice constants ($L>100$), and more than $10^7$ degrees of freedom (see Fig.~\ref{fig:surface_and_mirror_chern}). In contrast to previous theoretical estimations for the case of \alloy{}~\cite{Lent1986, Gao2008, Dziawa2012}, we find a critical concentration that matches the one found experimentally~\cite{Dimmock1966, Teo2008,Xu2012,Tanaka2012,Tanaka2013, Yan2014, Zhong2015}.

\co{Linear size of a disordered sample must be of order \texorpdfstring{$\xi$}{ξ}.}
\startsection{Review of existing algorithms}
We focus on the vicinity of disorder-driven phase transitions, where the localization length $\xi$ diverges.
The finite but potentially large value of $\xi$ defines the relevant length scale of our problem.
Two regions that are closer than $\xi$ feel each other's presence. Therefore, with open boundary conditions, the bulk has to be further than $\xi$ to the edge; analogously, with periodic boundary conditions when the linear size of the system is smaller than $\xi$ it hosts states whose extent is larger than the system size. These states span the whole system and overlap with themselves because of the finite size.
In order to simulate the bulk, the system needs to have a linear size $L\gtrsim\xi$.  Fluctuations of local quantities resulting from disorder scale with $\xi$, so averaging over a larger sample provides a good approximation of the thermodynamic limit.

\co{Most methods require full diagonalization and cost \texorpdfstring{$\xi^{3d}$}{ξ**(3d)}.}
The momentum-space formalism of topological invariants applies to disordered systems by studying a periodic system with a large disordered supercell of volume $\xi^d$.
This is equivalent to taking a finite torus and threading fluxes through its cycles~\cite{Avron1985, Niu1985, Essin2007}.
The final formula of the invariant is identical to the momentum space Berry curvature treatment applied to the supercell.
Other approaches include the Bott index~\cite{Loring2010}, topological markers~\cite{Bianco2011}, pseudospectra~\cite{Loring2015}, and noncommutative index theorems \cite{Prodan2013,Prodan2016,Akagi2017,Katsura2018}.
All of these methods involve diagonalization of a matrix of size proportional to the volume of the system.
Diagonalization scales as $N^3$ with the number of degrees of freedom $N$, so the computational cost of such methods is order $\xi^{3d}$, restricting them to small system sizes in three dimensions.

\co{Scattering approach is faster and requires \texorpdfstring{$\xi^{3d-3}$}{ξ**(3d-3)}.}
The scattering invariant formalism \cite{Fulga2012} avoids full diagonalization and only requires the knowledge of the scattering matrix at the Fermi level.
The most efficient known algorithm for computing the scattering matrix is based on the nested dissection method~\cite{George1973} and scales as $\xi^{3d - 3}$ for $d>1$.

\co{Projector can be approximated using KPM with a finite range of the expansion.}
\startsection{General strategy}
Computing the exact spectral projector
\begin{equation}
    \hat{P} = \theta(E_F - \hat{H}) = \sum_{n: E_n < E_F} \ket{n}\bra{n}
\end{equation}
by full diagonalization of the Hamiltonian is numerically expensive.
Instead, we approximate the projector using the kernel polynomial method with the Jackson kernel~\cite{Weisse2006}, detailed in Appendix~\ref{app:KPM}.
For a $d$-dimensional system of linear size $L$ the computational cost scales linearly with the number of degrees of freedom $L^d$, and with the number of moments $M$---the order of the expansion.
The order of the expansion sets a real-space cutoff in the approximate projector, which is an $M$'th order polynomial of the finite-range Hamiltonian.
In an insulating system, the projector is a local operator with matrix elements $\bra{\bm{x}}\hat{P}\ket{\bm{x'}} \propto \exp\left( -|\bm{x} - \bm{x'}| / \xi \right)$ that have a decay length $\xi$.
Hence, the error of the approximation scales as $\exp(- M/\xi)$, and the number of moments necessary for fixed precision scales linearly with the localization length as $M\sim\xi$~\cite{Aizenman1998, Prodan2005}. See Appendix~\ref{app:KPM}.

\co{Topological markers converge to the bulk invariant for a large subsystem.}
We use the topological marker formalism introduced by~\citet{Bianco2011}.
All $\ZZ$-valued topological markers are a partial trace per unit volume of a local operator~$\hat{\nu}$
\begin{equation}
\nu = \Tr_{S}(\hat{\nu}) = \frac{1}{|S|} \sum_{\lambda, \bm{x}\in S} \langle \bm{x}, \lambda | \hat{\nu} | \bm{x}, \lambda \rangle,
\end{equation}
where the sum runs over the sites $\bm{x}$ inside the subsystem $S$ with volume $|S|$, and their internal degrees of freedom $\lambda$.
The operator $\hat{\nu}$ is a polynomial of the spectral projector, position and symmetry operators, such that $\nu$ is dimensionless and independent of the detailed energetics or the overall length scale of the system.
The marker coincides with the momentum-space invariant in periodic systems, and converges to a quantized integer for large $S$ in insulating homogeneous disordered systems~\cite{Bianco2011}.

\co{We illustrate the general approach to \texorpdfstring{$Z$}{Z} invariants by considering Chern number, but we do specify how to do \texorpdfstring{$Z_2$}{Z2}.}
An example of topological marker is the real space expression for the Chern number~\cite{Bianco2011}:
\begin{equation}
    C = \Tr_{A} \hat{C} = 2 \pi \ii\Tr_{A} \left[\hat{P}\hat{x}\hat{P}, \hat{P}\hat{y}\hat{P} \right].
\end{equation}
Here, $A$ is the area of the subsystem, $\hat{x}$ and $\hat{y}$ are the two components of the position operator, and $[\cdot, \cdot]$ is the commutator.
Topological markers for all strong and weak $\ZZ$-valued topological invariants have similar algebraic expressions of the projected position operators~\cite{Song2014, MondragonShem2014, Chiu2016}, making a straightforward application of our method to these cases~\cite{in_preparation}.
We are not aware of similar formulations of $\mathbb{Z}_2$ topological indices suitable for KPM.

\co{To average the volume contribution to the invariant we apply stochastic trace.}
To estimate the trace per volume, we use the stochastic trace approximation~\cite{Weisse2006}
\begin{equation}
\Tr_S (\hat{\mathcal{O}}) \approx \frac{1}{R |S|} \sum_{i=1}^R \langle r_i | \hat{\mathcal{O}} | r_i \rangle,
\end{equation}
where $| r_i \rangle$ are random phase vectors localized in the region $S$.
The standard error of this approximation scales as $\sqrt{\xi^d/(R |S|)}$, meaning that the number of random vectors $R$ required for a given precision is constant if the system size is proportional to the localization length (see Appendix~\ref{app:stochastic_trace}).

\co{Repeating the supercell twice eliminates boundary fluctuations of the invariant.}
We build a supercell of size $L > \xi$. To ensure that the invariant $\nu$ obtained with the trace of the local marker converges to the momentum space topological invariant we must repeat the supercell with the disorder realization over the whole space. Since the approximation of the local operator $\hat{\nu}$ is localized with $\xi < L$, it is sufficient to repeat two times the supercell in every spatial direction, and compute the topological invariant as an average of the topological marker over the central $L^d$ volume. For a detailed description, see Appendix~\ref{app:geometry}.

\co{Combining all factors our method scales as \texorpdfstring{$\xi^{d+1}$}{ξ**(d+1)}.}
The resulting complexity of the computation depends linearly on the number of random vectors $R$ used (typically of order $1$), on the number of moments $M$, and on the number of sites of the system $L^d$.
We use a sparse representation of the short-ranged Hamiltonian. As a result the memory requirement scales linearly with the system size $L^d$ and is independent of other parameters.
Setting all quantities to their minimal values ($L\sim\xi$ and $M\sim\xi$), results in an algorithm with a computational cost scaling of $\xi^{d+1}$.

\co{Our method offers an edge in analysis of 3D materials, we therefore apply it to mirror Chern alloys.}%
\startsection{\label{sec:mirror_chern}Application to mirror Chern number}
Our method provides better scaling than existing approaches in $d \geq 3$.
We apply it to disordered 3D mirror Chern insulators.
These are a widely studied class of topological crystalline materials with a $\ZZ$ topological classification that relies on reflection symmetry~\cite{Hsieh2012}.
Several experimental realizations are known, including alloys~\cite{Xu2012, Dziawa2012, Tanaka2012,Tanaka2013,Yan2014, Zhong2015}.

\co{We rewrite the mirror Chern number in the topological marker formalism.}
In a reflection-symmetric system of fermions, all wave functions are eigenstates of the mirror operator $\hat{M}_z$, with eigenvalues $\pm \ii$.
The Chern numbers $C_\pm$ for mirror-even and mirror-odd wave functions are
\begin{equation}
    C_{\pm} = 2 \pi \ii \Tr_{A} \left[\tilde{x}_{\pm}, \tilde{y}_{\pm} \right],
\end{equation}
where ${\tilde{x}_{\pm} = \hat{M}_{\pm} \hat{P} \hat{x} \hat{P} \hat{M}_{\pm}}$, and ${\tilde{y}_{\pm} = \hat{M}_{\pm} \hat{P} \hat{y} \hat{P} \hat{M}_{\pm}}$ are the projected position operators restricted to the mirror-even or mirror-odd subspaces.
Here $\hat{M}_\pm$ are the projectors on the mirror-even and mirror-odd subspaces and $A$ is the area in the $xy$ plane.
The total Chern number $C$ is the sum of the Chern numbers for each subspace $C = C_+ + C_-$, while the mirror Chern number equals to their difference $C_M = \left( C_+ - C_-\right) / 2$.
In the presence of time-reversal invariance, the total Chern number vanishes (${C_+ = - C_-}$), and the mirror Chern number ${C_M = C_+}$ counts the helical surface modes.
Since the mirror operator ${\hat{M}_z = \ii \left(\hat{M}_+ - \hat{M}_-\right)}$ commutes with the projector $\hat{P}$ and position operators $\hat{x}$ and $\hat{y}$, we express the mirror Chern number as
\begin{equation}
    \label{eqn:mirror_chern}
    C_M = \pi \Tr_{A} \left(\hat{M}_z \left[\hat{P}\hat{x}\hat{P}, \hat{P}\hat{y}\hat{P} \right] \right).
\end{equation}

\co{In order to compute mirror Chern number we enforce reflection symmetry.}
In order to compute the mirror Chern number we consider a system with a disorder configuration that is mirror symmetric across the $M_1$ plane at $z=0$. The PBC in the $z$-direction results in another mirror plane $M_2$ across the boundary (see Appendix~\ref{app:geometry}).
The bulk of this system is locally indistinguishable from a sample without reflection symmetry except for the two mirror planes $M_1$ and $M_2$.
Therefore as long as the two mirror planes do not undergo a two dimensional topological transition, the mirror Chern number of the symmetric sample equals that of the bulk system\footnote{We expect that this will not happen with a regular short-range correlated disorder.}.

\co{\texorpdfstring{\alloy{}}{PbSnTe} is the canonical mirror Chern insulator.}
\emph{Tight binding model of \texorpdfstring{$\text{Pb}_{1-x}\text{Sn}_{x}\text{Te}$}{PbSnTe}.} ---
Topological crystalline insulators (TCI) protected by reflection symmetry \cite{Teo2008} were theoretically predicted~\cite{Hsieh2012} and experimentally observed~\cite{Tanaka2012, Dziawa2012, Xu2012,Tanaka2013,Yan2014, Zhong2015} in \alloy{} alloys.
They host metallic surface states on the surfaces that are symmetric with respect to the mirror plane~\cite{Hsieh2012,Fu2012,Chiu2016}.
Lead tin telluride was studied using either the virtual crystal approximation (VCA)~\cite{Lent1986, Dziawa2012} or \emph{ab initio} methods~\cite{Gao2008}, finding a gap closing and phase transition near $x=0.35$, or $x=0.23$, respectively. We use a tight-binding approach that captures long-range correlations, and find a different critical concentration.

\co{We consider substitutional disorder as appropriate for alloys.}
We consider the substitutional disorder of the lead tin telluride alloy \alloy{} coming from replacing some \ce{Pb} ions for \ce{Sn} ions.
This disorder is nonmagnetic, and it preserves the reflection symmetry on average, which is sufficient to protect the gapless surface states~\cite{Fulga2014, Ando2015}.
We disregard other types of symmetry breaking disorder appearing naturally in \alloy{} \cite{Ando2015}, such as ferroelectric structural distortion~\cite{Ando2015, Hsieh2012, Okada2013}, and magnetic dopants~\cite{Ando2015, Hsieh2012, Fang2014}.

\co{We use the full LCAO 18-orbital model and a simplified 6 orbital model.}
In our investigation we use two atomistic tight-binding models.
The first one includes 18 spinful $s$, $p$ and $d$ orbitals on both sublattices, with 36 bands in total.
This model accurately describes the energetics, using tight-binding parameters for both \ce{SnTe} and \ce{PbTe} derived from \emph{ab initio} simulations~\cite{Lent1986}.

To simulate the alloy, we substitute randomly \ce{Sn} for \ce{Te} with probability $x$.
We incorporate substitutional disorder by using the hopping amplitudes of \ce{SnTe} for \ce{Sn}--\ce{Te} bonds and \ce{PbTe} amplitudes  for \ce{Pb}--\ce{Te} bonds.
The onsite parameters of \ce{Te} atoms are slightly different in \ce{SnTe} and \ce{PbTe}; we therefore use a weighted average of these depending on the local environment.
We also use the appropriate onsite terms, including ${\bm{L} \cdot \bm{S}}$ spin-orbit coupling (SOC), depending on the type of the \ce{Sn} or \ce{Pb} atom.
Further details can be found in Appendix~\ref{app:18orb}.

When investigating the onsite energy dependent phase diagram of \ce{X_{1-x}Sn_{x}Te} alloys, we use a simplified model that only includes 6 spinful $p$ orbitals, with 12 bands \cite{Mitchell1966, Hsieh2012, Fulga2016}.
We include ${\bm{L} \cdot \bm{S}}$ SOC terms, first and second neighbor hoppings, with amplitudes that depend on the sublattices but not on the types of the atoms.
We restrict the effect of disorder to different onsite energies on \ce{Sn} and \ce{X} sites.
For more details, see Appendix~\ref{app:6orb}.

\co{We run heavy simulations using Kwant and KPM.}
\startsection{Results}
We define the Hamiltonians and perform the KPM expansions using the Kwant software package~\cite{Groth2014}.
The code to reproduce the figures in this article is available in Ref.~\cite{zenodo}.
First, we study the topological phase transition in the realistic 18-orbital model of \alloy{}.
We build a tight-binding model with PBC that preserves reflection symmetry and contains $W\times L_{1\overline{1}0} \times L_z$ unit cells, with 36 degrees of freedom each.
For the largest system size used this means \num{13824000} degrees of freedom in total.
This model accurately reproduces the energetics, resulting in full bandwidth of about \SI{25}{\electronvolt} and band gap of less than \SI{0.3}{\electronvolt}.
In order to resolve the gap that is multiple orders of magnitude smaller than the bandwidth, we use $M=\num{5000}$ moments in the calculation.
We use $R=5$ random vectors and 12 disorder realizations each.
This increases the time cost, but not the memory cost of the algorithm.

\begin{figure}[ht]
    \includegraphics[width=\columnwidth]{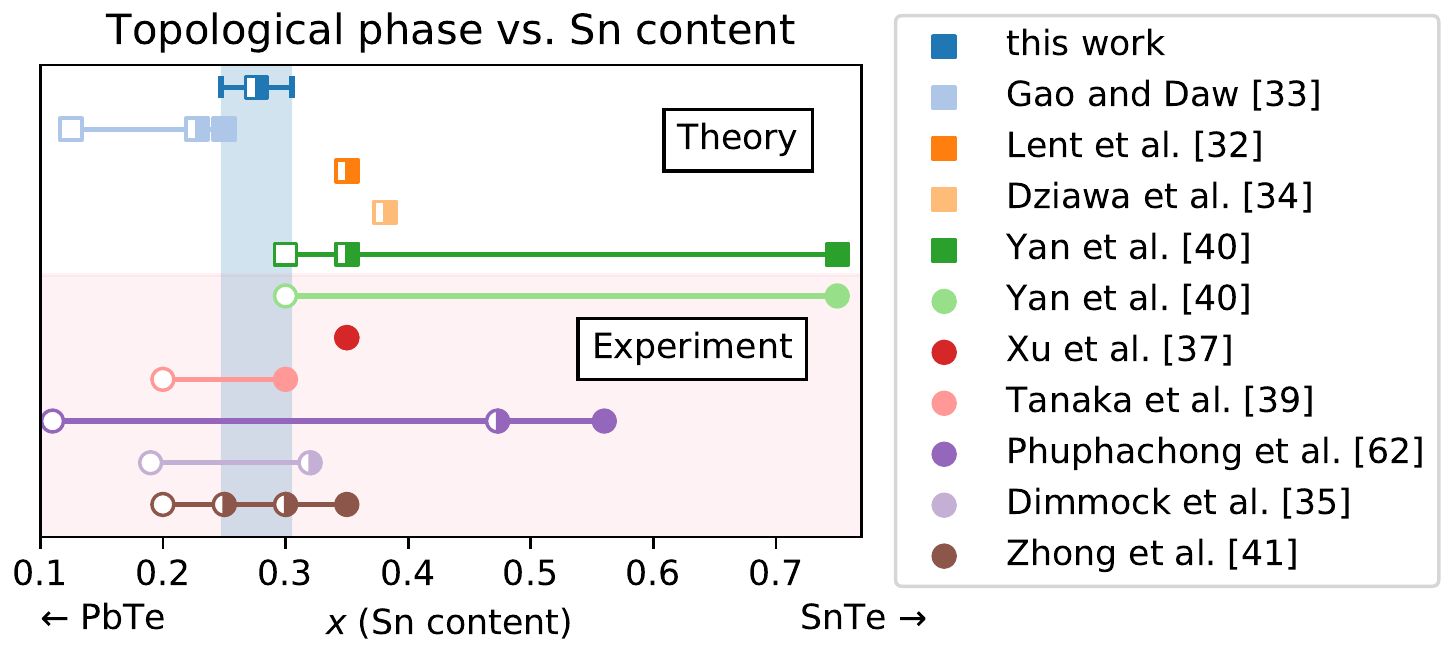}
    \caption{Comparison of the critical concentration obtained with our method, against theoretical approximations (squares) and experimental measurements (circles). The shapes are: empty for trivial phase, half-filled for estimated transition point (transition range for ~\citet{Zhong2015}), and fully filled for topological phase.}
    \label{fig:literature}
\end{figure}

\co{We find an accurate transition point.}
We perform finite size collapse of the data~\cite{Caio2019} (see Appendix~\ref{app:ffs}) and find the transition point at $x_c = \num{0.28(3)}$, with critical exponent $\nu=\num{0.9(6)}$ accurately describing the transition, see Fig.~\ref{fig:literature}.
This result improves significantly from the virtual crystal approximation (VCA) result~\cite{Lent1986, Dziawa2012, Yan2014}, and \emph{ab initio}~\cite{Gao2008}, but is consistent with the most precise experimental data~\cite{Tanaka2013, Zhong2015}. Other experimental estimates are made with only two values of the concentration $x$~\cite{Dimmock1966, Yan2014, Phuphachong2017}.

\begin{figure}[ht]
    \includegraphics[width=\columnwidth]{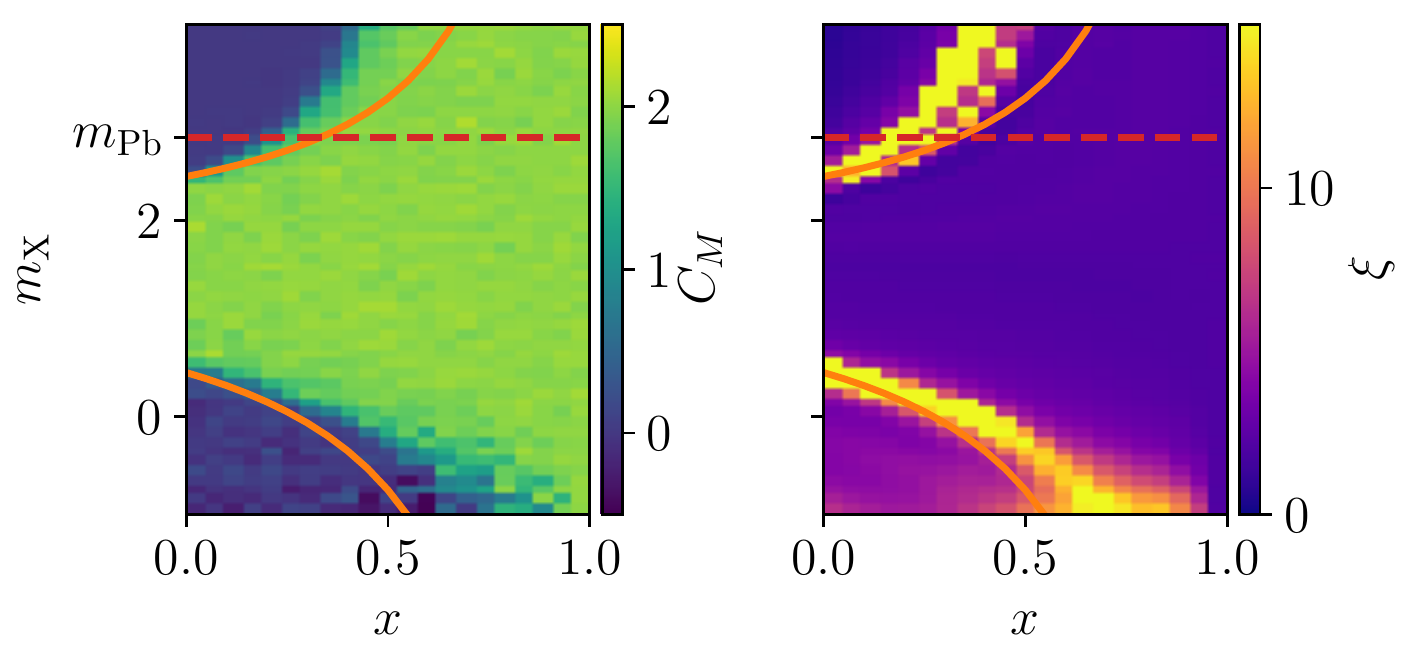}
    \caption{Mirror Chern number $C_M$ (left) and localization length $\xi$ (right) of \ce{X_{1-x}Sn_{x}Te} model as function of onsite energy $m_\text{X}$ and composition $x$.
    The overlay shows the phase boundary in the VCA and the horizontal dashed line corresponds to \ce{X}=\ce{Pb}.
    $\xi$ was calculated from the scaling of conductivity with sample sizes up to $L=20$ with 10 disorder realizations.
    $C_M$ was calculated in a system of ${\num{40} \times \num{40} \times \num{60}}$ unit cells, \num{2304000} degrees of freedom, $R=5$ random vectors and averaged over 4 disorder realizations.}
    \label{fig:3d_xi_CM}
\end{figure}

\co{With the 6-orbital model we find the phase diagram of a broader class of materials.}
To study a larger parameter space that includes other possible alloys of the \ce{X_{1-x}Sn_{x}Te} family that manifest the mirror Chern phase, we use the simplified 6-orbital model.
Besides the composition $x$, we also vary the onsite energy of the dopant cation $\ce{X}$, approximating compounds with lighter or heavier ions and similar electronic structures.
Figure~\ref{fig:3d_xi_CM} shows the phase diagram.
We find that the phase boundary differs from the VCA method, where the topological index only depends on the average onsite energy of the cations.

\co{We implemented the first practical approach to 3D topological invariants for alloys.}
\startsection{Conclusions}
Our method is the first to allow the computation of topological invariants of realistic 3D alloys.
Disorder in the crystalline structure is present in naturally found and artificially grown compounds, and it is inherent to substitutional alloys.
However, a computationally efficient method to analyze topological properties of realistic disordered materials was missing.

\co{Our method makes accurate prediction on the phase transition concentration.}
We apply our method to study the critical concentration of \alloy{}, and find an estimate that agrees with the most precise experimental data~\footnote{In the case of Ref.~\cite{Yan2014}, the ARPES data for $x=0.3$ is not conclusive, but it is interpreted with the help of VCA theory as a trivial phase. We claim the that for $x=0.3$, the experimental data and the theoretical results presented in~\cite{Yan2014} are not sufficient to ensure that $x=0.3$ lies on the trivial phase.}.
Whereas it is not possible to reconcile any of the other theoretical studies (known to us) with all the experimental data, specially Ref.~\cite{Zhong2015}.

\co{Our method has the best scaling.}
The scaling of computational time with system size of our method is better than the scaling of other methods available in the literature.
By using the kernel polynomial method, we achieve a computational time scaling of $\xi^{d+1}$ with the localization length.
Since we do not use eigenvalue solvers, only matrix-vector multiplication, the memory requirement only scales linearly with the sample volume as $\xi^d$.

\co{Our method is general and should allow automated search for topological alloys.}
Beyond Chern numbers, our formalism allows calculation of all $\ZZ$ valued strong and weak topological invariants in all dimensions~\cite{in_preparation}.
This method makes the automated discovery of topological alloys feasible, and can guide synthesis of new alloys in the future.
In this study we use energetically accurate tight-binding models obtained from \emph{ab initio} calculations performed on pure materials as input.
Using these tight-binding amplitudes for the atoms and bonds that appear in the alloy, we generate large disordered samples with various concentrations.
We are able to probe the topology, something that would not be accessible with other methods.
Simulation of small clusters with various disorder configurations is feasible using \emph{ab initio} methods~\cite{Gao2008}, tight-binding parameters obtained for all local environments would serve as a more accurate input for disordered models~\cite{Prodan2005}.

\co{Future research in \texorpdfstring{$\ZZ$}{Z} invariants, open questions on how to compute \texorpdfstring{$Z_2$}{Z2}, HOTI or topological semimetals.}
This work opens several directions for future research.
Our method is directly applicable to all types of disorder, as well as quasicrystalline and amorphous systems in symmetry classes that admit the topological marker formalism~\cite{in_preparation}.
This approach is not restricted to electrons in solids, and can be combined with finite element methods to analyze topology in disordered classical mechanical and photonic systems~\cite{Stutzer2018,Mitchell2018,Daraio2018,Zhou2018}.
We expect our method to perform well in disordered time-reversal breaking Weyl-semimetals with nonzero Hall conductivity, and further refinements could extend it to the time-reversal invariant case.
While we are not aware of a similar formulation of $\ZZ_2$ indices, KPM could be utilized to calculate quantized responses associated with these phases, such as the quantized magnetoelectric effect of 3D strong topological insulators~\cite{Bianco2013}.
A similar approach could also be applied to higher order topological insulators to calculate multipole moments of the charge density~\cite{Benalcazar2017}.
We expect that KPM could be used to study a wide variety of related topics in condensed matter physics, such as probing localization, topological Anderson insulators, or numerical renormalization group studies of the topological markers.

\begin{acknowledgments}
\startsection{Author contributions}
The project to study topological invariants using KPM was initiated by P. Perez-Piskunow and D. Varjas, the scope of the project was later refined using contributions from all authors.
P. Perez-Piskunow and D. Varjas wrote the code used for the numerical calculations and performed calculations.
D. Varjas performed the large-scale numerical calculations on the \alloy{} tight-binding models.
All authors took part in the analysis of the method and writing the manuscript.

\startsection{Acknowledgments}
We are grateful to M. D. Caio for discussions and insights on local topological markers, M. Hoskam for help with implementing the tight-binding models, and B. Nijholt for technical assistance with the numerical calculations.
We thank S. Roche and A. Cummings for critical revision of the manuscript, and P. Brouwer for helpful discussions.
This work was supported by ERC Starting Grant 638760, the Netherlands Organisation for Scientific Research (NWO/OCW) as part of the Frontiers of Nanoscience program, NWO VIDI grant 680-47-53, the US Office of Naval Research, University of Chicago MRSEC through a Kadanoff-Rice postdoctoral fellowship, and the European Union’s Horizon 2020 research and innovation programme under grant agreement No 824140.
\end{acknowledgments}

\bibliography{bibliography}

\begin{thebibliography}{72}%
\makeatletter
\providecommand \@ifxundefined [1]{%
 \@ifx{#1\undefined}
}%
\providecommand \@ifnum [1]{%
 \ifnum #1\expandafter \@firstoftwo
 \else \expandafter \@secondoftwo
 \fi
}%
\providecommand \@ifx [1]{%
 \ifx #1\expandafter \@firstoftwo
 \else \expandafter \@secondoftwo
 \fi
}%
\providecommand \natexlab [1]{#1}%
\providecommand \enquote  [1]{``#1''}%
\providecommand \bibnamefont  [1]{#1}%
\providecommand \bibfnamefont [1]{#1}%
\providecommand \citenamefont [1]{#1}%
\providecommand \href@noop [0]{\@secondoftwo}%
\providecommand \href [0]{\begingroup \@sanitize@url \@href}%
\providecommand \@href[1]{\@@startlink{#1}\@@href}%
\providecommand \@@href[1]{\endgroup#1\@@endlink}%
\providecommand \@sanitize@url [0]{\catcode `\\12\catcode `\$12\catcode
  `\&12\catcode `\#12\catcode `\^12\catcode `\_12\catcode `\%12\relax}%
\providecommand \@@startlink[1]{}%
\providecommand \@@endlink[0]{}%
\providecommand \url  [0]{\begingroup\@sanitize@url \@url }%
\providecommand \@url [1]{\endgroup\@href {#1}{\urlprefix }}%
\providecommand \urlprefix  [0]{URL }%
\providecommand \Eprint [0]{\href }%
\providecommand \doibase [0]{http://dx.doi.org/}%
\providecommand \selectlanguage [0]{\@gobble}%
\providecommand \bibinfo  [0]{\@secondoftwo}%
\providecommand \bibfield  [0]{\@secondoftwo}%
\providecommand \translation [1]{[#1]}%
\providecommand \BibitemOpen [0]{}%
\providecommand \bibitemStop [0]{}%
\providecommand \bibitemNoStop [0]{.\EOS\space}%
\providecommand \EOS [0]{\spacefactor3000\relax}%
\providecommand \BibitemShut  [1]{\csname bibitem#1\endcsname}%
\let\auto@bib@innerbib\@empty
\bibitem [{\citenamefont {Qi}\ and\ \citenamefont
  {Zhang}(2011)}]{qi2011topological}%
  \BibitemOpen
  \bibfield  {author} {\bibinfo {author} {\bibfnamefont {X.-L.}\ \bibnamefont
  {Qi}}\ and\ \bibinfo {author} {\bibfnamefont {S.-C.}\ \bibnamefont {Zhang}},\
  }\href {https://link.aps.org/doi/10.1103/RevModPhys.83.1057} {\bibfield
  {journal} {\bibinfo  {journal} {Rev. Mod. Phys.}\ }\textbf {\bibinfo {volume}
  {83}},\ \bibinfo {pages} {1057} (\bibinfo {year} {2011})}\BibitemShut
  {NoStop}%
\bibitem [{\citenamefont {Chiu}\ \emph {et~al.}(2016)\citenamefont {Chiu},
  \citenamefont {Teo}, \citenamefont {Schnyder},\ and\ \citenamefont
  {Ryu}}]{Chiu2016}%
  \BibitemOpen
  \bibfield  {author} {\bibinfo {author} {\bibfnamefont {C.-K.}\ \bibnamefont
  {Chiu}}, \bibinfo {author} {\bibfnamefont {J.~C.~Y.}\ \bibnamefont {Teo}},
  \bibinfo {author} {\bibfnamefont {A.~P.}\ \bibnamefont {Schnyder}}, \ and\
  \bibinfo {author} {\bibfnamefont {S.}~\bibnamefont {Ryu}},\ }\href {\doibase
  10.1103/revmodphys.88.035005} {\bibfield  {journal} {\bibinfo  {journal}
  {Reviews of Modern Physics}\ }\textbf {\bibinfo {volume} {88}},\ \bibinfo
  {pages} {035005} (\bibinfo {year} {2016})}\BibitemShut {NoStop}%
\bibitem [{\citenamefont {Kruthoff}\ \emph {et~al.}(2017)\citenamefont
  {Kruthoff}, \citenamefont {de~Boer}, \citenamefont {van Wezel}, \citenamefont
  {Kane},\ and\ \citenamefont {Slager}}]{Kruthoff2017}%
  \BibitemOpen
  \bibfield  {author} {\bibinfo {author} {\bibfnamefont {J.}~\bibnamefont
  {Kruthoff}}, \bibinfo {author} {\bibfnamefont {J.}~\bibnamefont {de~Boer}},
  \bibinfo {author} {\bibfnamefont {J.}~\bibnamefont {van Wezel}}, \bibinfo
  {author} {\bibfnamefont {C.~L.}\ \bibnamefont {Kane}}, \ and\ \bibinfo
  {author} {\bibfnamefont {R.-J.}\ \bibnamefont {Slager}},\ }\href {\doibase
  10.1103/PhysRevX.7.041069} {\bibfield  {journal} {\bibinfo  {journal} {Phys.
  Rev. X}\ }\textbf {\bibinfo {volume} {7}},\ \bibinfo {pages} {041069}
  (\bibinfo {year} {2017})}\BibitemShut {NoStop}%
\bibitem [{\citenamefont {Po}\ \emph {et~al.}(2017)\citenamefont {Po},
  \citenamefont {Vishwanath},\ and\ \citenamefont
  {Watanabe}}]{po2017classification}%
  \BibitemOpen
  \bibfield  {author} {\bibinfo {author} {\bibfnamefont {H.~C.}\ \bibnamefont
  {Po}}, \bibinfo {author} {\bibfnamefont {A.}~\bibnamefont {Vishwanath}}, \
  and\ \bibinfo {author} {\bibfnamefont {H.}~\bibnamefont {Watanabe}},\ }\href
  {\doibase 10.1038/s41467-017-00133-2} {\bibfield  {journal} {\bibinfo
  {journal} {Nat. Commun.}\ }\textbf {\bibinfo {volume} {8}},\  (\bibinfo
  {year} {2017})}\BibitemShut {NoStop}%
\bibitem [{\citenamefont {{Bradlyn}}\ \emph {et~al.}(2017)\citenamefont
  {{Bradlyn}}, \citenamefont {{Elcoro}}, \citenamefont {{Cano}}, \citenamefont
  {{Vergniory}}, \citenamefont {{Wang}}, \citenamefont {{Felser}},
  \citenamefont {{Aroyo}},\ and\ \citenamefont {{Bernevig}}}]{Bradlyn2017}%
  \BibitemOpen
  \bibfield  {author} {\bibinfo {author} {\bibfnamefont {B.}~\bibnamefont
  {{Bradlyn}}}, \bibinfo {author} {\bibfnamefont {L.}~\bibnamefont {{Elcoro}}},
  \bibinfo {author} {\bibfnamefont {J.}~\bibnamefont {{Cano}}}, \bibinfo
  {author} {\bibfnamefont {M.~G.}\ \bibnamefont {{Vergniory}}}, \bibinfo
  {author} {\bibfnamefont {Z.}~\bibnamefont {{Wang}}}, \bibinfo {author}
  {\bibfnamefont {C.}~\bibnamefont {{Felser}}}, \bibinfo {author}
  {\bibfnamefont {M.~I.}\ \bibnamefont {{Aroyo}}}, \ and\ \bibinfo {author}
  {\bibfnamefont {B.~A.}\ \bibnamefont {{Bernevig}}},\ }\href {\doibase
  10.1038/nature23268} {\bibfield  {journal} {\bibinfo  {journal} {Nature}\
  }\textbf {\bibinfo {volume} {547}},\ \bibinfo {pages} {298} (\bibinfo {year}
  {2017})}\BibitemShut {NoStop}%
\bibitem [{\citenamefont {Yang}\ \emph {et~al.}(2012)\citenamefont {Yang},
  \citenamefont {Setyawan}, \citenamefont {Wang}, \citenamefont {Nardelli},\
  and\ \citenamefont {Curtarolo}}]{Yang2012}%
  \BibitemOpen
  \bibfield  {author} {\bibinfo {author} {\bibfnamefont {K.}~\bibnamefont
  {Yang}}, \bibinfo {author} {\bibfnamefont {W.}~\bibnamefont {Setyawan}},
  \bibinfo {author} {\bibfnamefont {S.}~\bibnamefont {Wang}}, \bibinfo {author}
  {\bibfnamefont {M.~B.}\ \bibnamefont {Nardelli}}, \ and\ \bibinfo {author}
  {\bibfnamefont {S.}~\bibnamefont {Curtarolo}},\ }\href {\doibase
  10.1038/nmat3332} {\bibfield  {journal} {\bibinfo  {journal} {Nature
  Materials}\ }\textbf {\bibinfo {volume} {11}},\ \bibinfo {pages} {614}
  (\bibinfo {year} {2012})}\BibitemShut {NoStop}%
\bibitem [{\citenamefont {Curtarolo}\ \emph {et~al.}(2013)\citenamefont
  {Curtarolo}, \citenamefont {Hart}, \citenamefont {Nardelli}, \citenamefont
  {Mingo}, \citenamefont {Sanvito},\ and\ \citenamefont
  {Levy}}]{Curtarolo2013}%
  \BibitemOpen
  \bibfield  {author} {\bibinfo {author} {\bibfnamefont {S.}~\bibnamefont
  {Curtarolo}}, \bibinfo {author} {\bibfnamefont {G.~L.~W.}\ \bibnamefont
  {Hart}}, \bibinfo {author} {\bibfnamefont {M.~B.}\ \bibnamefont {Nardelli}},
  \bibinfo {author} {\bibfnamefont {N.}~\bibnamefont {Mingo}}, \bibinfo
  {author} {\bibfnamefont {S.}~\bibnamefont {Sanvito}}, \ and\ \bibinfo
  {author} {\bibfnamefont {O.}~\bibnamefont {Levy}},\ }\href {\doibase
  10.1038/nmat3568} {\bibfield  {journal} {\bibinfo  {journal} {Nature
  Materials}\ }\textbf {\bibinfo {volume} {12}},\ \bibinfo {pages} {191}
  (\bibinfo {year} {2013})}\BibitemShut {NoStop}%
\bibitem [{\citenamefont {Zhang}\ \emph {et~al.}(2018)\citenamefont {Zhang},
  \citenamefont {Zhang}, \citenamefont {Li}, \citenamefont {Koepernik},
  \citenamefont {Yao},\ and\ \citenamefont {Zhang}}]{Zhang2018b}%
  \BibitemOpen
  \bibfield  {author} {\bibinfo {author} {\bibfnamefont {Z.}~\bibnamefont
  {Zhang}}, \bibinfo {author} {\bibfnamefont {R.-W.}\ \bibnamefont {Zhang}},
  \bibinfo {author} {\bibfnamefont {X.}~\bibnamefont {Li}}, \bibinfo {author}
  {\bibfnamefont {K.}~\bibnamefont {Koepernik}}, \bibinfo {author}
  {\bibfnamefont {Y.}~\bibnamefont {Yao}}, \ and\ \bibinfo {author}
  {\bibfnamefont {H.}~\bibnamefont {Zhang}},\ }\href {\doibase
  10.1021/acs.jpclett.8b02800} {\bibfield  {journal} {\bibinfo  {journal} {The
  Journal of Physical Chemistry Letters}\ ,\ \bibinfo {pages} {6224}} (\bibinfo
  {year} {2018})}\BibitemShut {NoStop}%
\bibitem [{\citenamefont {Tang}\ \emph
  {et~al.}(2019{\natexlab{a}})\citenamefont {Tang}, \citenamefont {Po},
  \citenamefont {Vishwanath},\ and\ \citenamefont {Wan}}]{Tang2018a}%
  \BibitemOpen
  \bibfield  {author} {\bibinfo {author} {\bibfnamefont {F.}~\bibnamefont
  {Tang}}, \bibinfo {author} {\bibfnamefont {H.~C.}\ \bibnamefont {Po}},
  \bibinfo {author} {\bibfnamefont {A.}~\bibnamefont {Vishwanath}}, \ and\
  \bibinfo {author} {\bibfnamefont {X.}~\bibnamefont {Wan}},\ }\href {\doibase
  10.1038/s41586-019-0937-5} {\bibfield  {journal} {\bibinfo  {journal}
  {Nature}\ }\textbf {\bibinfo {volume} {566}},\ \bibinfo {pages} {486}
  (\bibinfo {year} {2019}{\natexlab{a}})}\BibitemShut {NoStop}%
\bibitem [{\citenamefont {Tang}\ \emph
  {et~al.}(2019{\natexlab{b}})\citenamefont {Tang}, \citenamefont {Po},
  \citenamefont {Vishwanath},\ and\ \citenamefont {Wan}}]{Tang2018b}%
  \BibitemOpen
  \bibfield  {author} {\bibinfo {author} {\bibfnamefont {F.}~\bibnamefont
  {Tang}}, \bibinfo {author} {\bibfnamefont {H.~C.}\ \bibnamefont {Po}},
  \bibinfo {author} {\bibfnamefont {A.}~\bibnamefont {Vishwanath}}, \ and\
  \bibinfo {author} {\bibfnamefont {X.}~\bibnamefont {Wan}},\ }\href {\doibase
  10.1038/s41567-019-0418-7} {\bibfield  {journal} {\bibinfo  {journal} {Nature
  Physics}\ }\textbf {\bibinfo {volume} {15}},\ \bibinfo {pages} {470}
  (\bibinfo {year} {2019}{\natexlab{b}})}\BibitemShut {NoStop}%
\bibitem [{\citenamefont {Zhang}\ \emph {et~al.}(2019)\citenamefont {Zhang},
  \citenamefont {Jiang}, \citenamefont {Song}, \citenamefont {Huang},
  \citenamefont {He}, \citenamefont {Fang}, \citenamefont {Weng},\ and\
  \citenamefont {Fang}}]{Zhang2018}%
  \BibitemOpen
  \bibfield  {author} {\bibinfo {author} {\bibfnamefont {T.}~\bibnamefont
  {Zhang}}, \bibinfo {author} {\bibfnamefont {Y.}~\bibnamefont {Jiang}},
  \bibinfo {author} {\bibfnamefont {Z.}~\bibnamefont {Song}}, \bibinfo {author}
  {\bibfnamefont {H.}~\bibnamefont {Huang}}, \bibinfo {author} {\bibfnamefont
  {Y.}~\bibnamefont {He}}, \bibinfo {author} {\bibfnamefont {Z.}~\bibnamefont
  {Fang}}, \bibinfo {author} {\bibfnamefont {H.}~\bibnamefont {Weng}}, \ and\
  \bibinfo {author} {\bibfnamefont {C.}~\bibnamefont {Fang}},\ }\href {\doibase
  10.1038/s41586-019-0944-6} {\bibfield  {journal} {\bibinfo  {journal}
  {Nature}\ }\textbf {\bibinfo {volume} {566}},\ \bibinfo {pages} {475}
  (\bibinfo {year} {2019})}\BibitemShut {NoStop}%
\bibitem [{\citenamefont {Vergniory}\ \emph {et~al.}(2019)\citenamefont
  {Vergniory}, \citenamefont {Elcoro}, \citenamefont {Felser}, \citenamefont
  {Regnault}, \citenamefont {Bernevig},\ and\ \citenamefont
  {Wang}}]{Vergniory2018}%
  \BibitemOpen
  \bibfield  {author} {\bibinfo {author} {\bibfnamefont {M.~G.}\ \bibnamefont
  {Vergniory}}, \bibinfo {author} {\bibfnamefont {L.}~\bibnamefont {Elcoro}},
  \bibinfo {author} {\bibfnamefont {C.}~\bibnamefont {Felser}}, \bibinfo
  {author} {\bibfnamefont {N.}~\bibnamefont {Regnault}}, \bibinfo {author}
  {\bibfnamefont {B.~A.}\ \bibnamefont {Bernevig}}, \ and\ \bibinfo {author}
  {\bibfnamefont {Z.}~\bibnamefont {Wang}},\ }\href {\doibase
  10.1038/s41586-019-0954-4} {\bibfield  {journal} {\bibinfo  {journal}
  {Nature}\ }\textbf {\bibinfo {volume} {566}},\ \bibinfo {pages} {480}
  (\bibinfo {year} {2019})}\BibitemShut {NoStop}%
\bibitem [{\citenamefont {Fu}\ and\ \citenamefont {Kane}(2007)}]{Fu2007}%
  \BibitemOpen
  \bibfield  {author} {\bibinfo {author} {\bibfnamefont {L.}~\bibnamefont
  {Fu}}\ and\ \bibinfo {author} {\bibfnamefont {C.~L.}\ \bibnamefont {Kane}},\
  }\href {\doibase 10.1103/physrevb.76.045302} {\bibfield  {journal} {\bibinfo
  {journal} {Physical Review B}\ }\textbf {\bibinfo {volume} {76}},\ \bibinfo
  {pages} {045302} (\bibinfo {year} {2007})}\BibitemShut {NoStop}%
\bibitem [{\citenamefont {Hsieh}\ \emph {et~al.}(2008)\citenamefont {Hsieh},
  \citenamefont {Qian}, \citenamefont {Wray}, \citenamefont {Xia},
  \citenamefont {Hor}, \citenamefont {Cava},\ and\ \citenamefont
  {Hasan}}]{Hsieh2008}%
  \BibitemOpen
  \bibfield  {author} {\bibinfo {author} {\bibfnamefont {D.}~\bibnamefont
  {Hsieh}}, \bibinfo {author} {\bibfnamefont {D.}~\bibnamefont {Qian}},
  \bibinfo {author} {\bibfnamefont {L.}~\bibnamefont {Wray}}, \bibinfo {author}
  {\bibfnamefont {Y.}~\bibnamefont {Xia}}, \bibinfo {author} {\bibfnamefont
  {Y.~S.}\ \bibnamefont {Hor}}, \bibinfo {author} {\bibfnamefont {R.~J.}\
  \bibnamefont {Cava}}, \ and\ \bibinfo {author} {\bibfnamefont {M.~Z.}\
  \bibnamefont {Hasan}},\ }\href {\doibase 10.1038/nature06843} {\bibfield
  {journal} {\bibinfo  {journal} {Nature}\ }\textbf {\bibinfo {volume} {452}},\
  \bibinfo {pages} {970} (\bibinfo {year} {2008})}\BibitemShut {NoStop}%
\bibitem [{\citenamefont {Chadov}\ \emph {et~al.}(2013)\citenamefont {Chadov},
  \citenamefont {Kiss}, \citenamefont {Kübler},\ and\ \citenamefont
  {Felser}}]{Chadov2013}%
  \BibitemOpen
  \bibfield  {author} {\bibinfo {author} {\bibfnamefont {S.}~\bibnamefont
  {Chadov}}, \bibinfo {author} {\bibfnamefont {J.}~\bibnamefont {Kiss}},
  \bibinfo {author} {\bibfnamefont {J.}~\bibnamefont {Kübler}}, \ and\
  \bibinfo {author} {\bibfnamefont {C.}~\bibnamefont {Felser}},\ }\href
  {\doibase 10.1002/pssr.201206395} {\bibfield  {journal} {\bibinfo  {journal}
  {physica status solidi (RRL) – Rapid Research Letters}\ }\textbf {\bibinfo
  {volume} {7}},\ \bibinfo {pages} {82} (\bibinfo {year} {2013})}\BibitemShut
  {NoStop}%
\bibitem [{\citenamefont {Sante}\ \emph {et~al.}(2015)\citenamefont {Sante},
  \citenamefont {Barone}, \citenamefont {Plekhanov}, \citenamefont {Ciuchi},\
  and\ \citenamefont {Picozzi}}]{Sante2015}%
  \BibitemOpen
  \bibfield  {author} {\bibinfo {author} {\bibfnamefont {D.}~\bibnamefont
  {Sante}}, \bibinfo {author} {\bibfnamefont {P.}~\bibnamefont {Barone}},
  \bibinfo {author} {\bibfnamefont {E.}~\bibnamefont {Plekhanov}}, \bibinfo
  {author} {\bibfnamefont {S.}~\bibnamefont {Ciuchi}}, \ and\ \bibinfo {author}
  {\bibfnamefont {S.}~\bibnamefont {Picozzi}},\ }\href {\doibase
  10.1038/srep11285} {\bibfield  {journal} {\bibinfo  {journal} {Scientific
  reports}\ }\textbf {\bibinfo {volume} {5}},\ \bibinfo {pages} {11285}
  (\bibinfo {year} {2015})}\BibitemShut {NoStop}%
\bibitem [{\citenamefont {Li}\ \emph {et~al.}(2009)\citenamefont {Li},
  \citenamefont {Chu}, \citenamefont {Jain},\ and\ \citenamefont
  {Shen}}]{Li2009}%
  \BibitemOpen
  \bibfield  {author} {\bibinfo {author} {\bibfnamefont {J.}~\bibnamefont
  {Li}}, \bibinfo {author} {\bibfnamefont {R.-L.}\ \bibnamefont {Chu}},
  \bibinfo {author} {\bibfnamefont {J.~K.}\ \bibnamefont {Jain}}, \ and\
  \bibinfo {author} {\bibfnamefont {S.-Q.}\ \bibnamefont {Shen}},\ }\href
  {\doibase 10.1103/PhysRevLett.102.136806} {\bibfield  {journal} {\bibinfo
  {journal} {Phys. Rev. Lett.}\ }\textbf {\bibinfo {volume} {102}},\ \bibinfo
  {pages} {136806} (\bibinfo {year} {2009})}\BibitemShut {NoStop}%
\bibitem [{\citenamefont {Groth}\ \emph {et~al.}(2009)\citenamefont {Groth},
  \citenamefont {Wimmer}, \citenamefont {Akhmerov}, \citenamefont
  {Tworzyd\l{}o},\ and\ \citenamefont {Beenakker}}]{Groth2009}%
  \BibitemOpen
  \bibfield  {author} {\bibinfo {author} {\bibfnamefont {C.~W.}\ \bibnamefont
  {Groth}}, \bibinfo {author} {\bibfnamefont {M.}~\bibnamefont {Wimmer}},
  \bibinfo {author} {\bibfnamefont {A.~R.}\ \bibnamefont {Akhmerov}}, \bibinfo
  {author} {\bibfnamefont {J.}~\bibnamefont {Tworzyd\l{}o}}, \ and\ \bibinfo
  {author} {\bibfnamefont {C.~W.~J.}\ \bibnamefont {Beenakker}},\ }\href
  {\doibase 10.1103/PhysRevLett.103.196805} {\bibfield  {journal} {\bibinfo
  {journal} {Phys. Rev. Lett.}\ }\textbf {\bibinfo {volume} {103}},\ \bibinfo
  {pages} {196805} (\bibinfo {year} {2009})}\BibitemShut {NoStop}%
\bibitem [{\citenamefont {Kitaev}(2006)}]{Kitaev2006}%
  \BibitemOpen
  \bibfield  {author} {\bibinfo {author} {\bibfnamefont {A.}~\bibnamefont
  {Kitaev}},\ }\href {\doibase 10.1016/j.aop.2005.10.005} {\bibfield  {journal}
  {\bibinfo  {journal} {Annals of Physics}\ }\textbf {\bibinfo {volume}
  {321}},\ \bibinfo {pages} {2} (\bibinfo {year} {2006})}\BibitemShut {NoStop}%
\bibitem [{\citenamefont {Essin}\ and\ \citenamefont
  {Moore}(2007)}]{Essin2007}%
  \BibitemOpen
  \bibfield  {author} {\bibinfo {author} {\bibfnamefont {A.~M.}\ \bibnamefont
  {Essin}}\ and\ \bibinfo {author} {\bibfnamefont {J.~E.}\ \bibnamefont
  {Moore}},\ }\href@noop {} {\bibfield  {journal} {\bibinfo  {journal}
  {Physical Review B}\ }\textbf {\bibinfo {volume} {76}},\ \bibinfo {pages}
  {165307} (\bibinfo {year} {2007})}\BibitemShut {NoStop}%
\bibitem [{\citenamefont {Loring}\ and\ \citenamefont
  {Hastings}(2010)}]{Loring2010}%
  \BibitemOpen
  \bibfield  {author} {\bibinfo {author} {\bibfnamefont {T.~A.}\ \bibnamefont
  {Loring}}\ and\ \bibinfo {author} {\bibfnamefont {M.~B.}\ \bibnamefont
  {Hastings}},\ }\href {\doibase 10.1209/0295-5075/92/67004} {\bibfield
  {journal} {\bibinfo  {journal} {{EPL} (Europhysics Letters)}\ }\textbf
  {\bibinfo {volume} {92}},\ \bibinfo {pages} {67004} (\bibinfo {year}
  {2010})}\BibitemShut {NoStop}%
\bibitem [{\citenamefont {Bianco}\ and\ \citenamefont
  {Resta}(2011)}]{Bianco2011}%
  \BibitemOpen
  \bibfield  {author} {\bibinfo {author} {\bibfnamefont {R.}~\bibnamefont
  {Bianco}}\ and\ \bibinfo {author} {\bibfnamefont {R.}~\bibnamefont {Resta}},\
  }\href {\doibase 10.1103/physrevb.84.241106} {\bibfield  {journal} {\bibinfo
  {journal} {Physical Review B}\ }\textbf {\bibinfo {volume} {84}},\ \bibinfo
  {pages} {241106(R)} (\bibinfo {year} {2011})}\BibitemShut {NoStop}%
\bibitem [{\citenamefont {Loring}(2015)}]{Loring2015}%
  \BibitemOpen
  \bibfield  {author} {\bibinfo {author} {\bibfnamefont {T.~A.}\ \bibnamefont
  {Loring}},\ }\href {\doibase https://doi.org/10.1016/j.aop.2015.02.031}
  {\bibfield  {journal} {\bibinfo  {journal} {Annals of Physics}\ }\textbf
  {\bibinfo {volume} {356}},\ \bibinfo {pages} {383 } (\bibinfo {year}
  {2015})}\BibitemShut {NoStop}%
\bibitem [{\citenamefont {Prodan}(2013)}]{Prodan2013}%
  \BibitemOpen
  \bibfield  {author} {\bibinfo {author} {\bibfnamefont {E.}~\bibnamefont
  {Prodan}},\ }\href {\doibase 10.1093/amrx/abs017} {\bibfield  {journal}
  {\bibinfo  {journal} {Applied Mathematics Research eXpress}\ }\textbf
  {\bibinfo {volume} {2013}},\ \bibinfo {pages} {176} (\bibinfo {year}
  {2013})}\BibitemShut {NoStop}%
\bibitem [{\citenamefont {Prodan}\ and\ \citenamefont
  {Schulz-Baldes}(2016)}]{Prodan2016}%
  \BibitemOpen
  \bibfield  {author} {\bibinfo {author} {\bibfnamefont {E.}~\bibnamefont
  {Prodan}}\ and\ \bibinfo {author} {\bibfnamefont {H.}~\bibnamefont
  {Schulz-Baldes}},\ }\href {\doibase 10.1007/978-3-319-29351-6} {\emph
  {\bibinfo {title} {Bulk and Boundary Invariants for Complex Topological
  Insulators}}}\ (\bibinfo  {publisher} {Springer International Publishing},\
  \bibinfo {year} {2016})\BibitemShut {NoStop}%
\bibitem [{\citenamefont {Akagi}\ \emph {et~al.}(2017)\citenamefont {Akagi},
  \citenamefont {Katsura},\ and\ \citenamefont {Koma}}]{Akagi2017}%
  \BibitemOpen
  \bibfield  {author} {\bibinfo {author} {\bibfnamefont {Y.}~\bibnamefont
  {Akagi}}, \bibinfo {author} {\bibfnamefont {H.}~\bibnamefont {Katsura}}, \
  and\ \bibinfo {author} {\bibfnamefont {T.}~\bibnamefont {Koma}},\ }\href
  {\doibase 10.7566/JPSJ.86.123710} {\bibfield  {journal} {\bibinfo  {journal}
  {Journal of the Physical Society of Japan}\ }\textbf {\bibinfo {volume}
  {86}},\ \bibinfo {pages} {123710} (\bibinfo {year} {2017})}\BibitemShut
  {NoStop}%
\bibitem [{\citenamefont {Katsura}\ and\ \citenamefont
  {Koma}(2018)}]{Katsura2018}%
  \BibitemOpen
  \bibfield  {author} {\bibinfo {author} {\bibfnamefont {H.}~\bibnamefont
  {Katsura}}\ and\ \bibinfo {author} {\bibfnamefont {T.}~\bibnamefont {Koma}},\
  }\href {\doibase 10.1063/1.5026964} {\bibfield  {journal} {\bibinfo
  {journal} {Journal of Mathematical Physics}\ }\textbf {\bibinfo {volume}
  {59}},\ \bibinfo {pages} {031903} (\bibinfo {year} {2018})}\BibitemShut
  {NoStop}%
\bibitem [{\citenamefont {Fulga}\ \emph {et~al.}(2014)\citenamefont {Fulga},
  \citenamefont {van Heck}, \citenamefont {Edge},\ and\ \citenamefont
  {Akhmerov}}]{Fulga2014}%
  \BibitemOpen
  \bibfield  {author} {\bibinfo {author} {\bibfnamefont {I.~C.}\ \bibnamefont
  {Fulga}}, \bibinfo {author} {\bibfnamefont {B.}~\bibnamefont {van Heck}},
  \bibinfo {author} {\bibfnamefont {J.~M.}\ \bibnamefont {Edge}}, \ and\
  \bibinfo {author} {\bibfnamefont {A.~R.}\ \bibnamefont {Akhmerov}},\ }\href
  {\doibase 10.1103/physrevb.89.155424} {\bibfield  {journal} {\bibinfo
  {journal} {Physical Review B}\ }\textbf {\bibinfo {volume} {89}},\ \bibinfo
  {pages} {155424} (\bibinfo {year} {2014})}\BibitemShut {NoStop}%
\bibitem [{\citenamefont {Wei{\ss}e}\ \emph {et~al.}(2006)\citenamefont
  {Wei{\ss}e}, \citenamefont {Wellein}, \citenamefont {Alvermann},\ and\
  \citenamefont {Fehske}}]{Weisse2006}%
  \BibitemOpen
  \bibfield  {author} {\bibinfo {author} {\bibfnamefont {A.}~\bibnamefont
  {Wei{\ss}e}}, \bibinfo {author} {\bibfnamefont {G.}~\bibnamefont {Wellein}},
  \bibinfo {author} {\bibfnamefont {A.}~\bibnamefont {Alvermann}}, \ and\
  \bibinfo {author} {\bibfnamefont {H.}~\bibnamefont {Fehske}},\ }\href
  {\doibase 10.1103/revmodphys.78.275} {\bibfield  {journal} {\bibinfo
  {journal} {Reviews of Modern Physics}\ }\textbf {\bibinfo {volume} {78}},\
  \bibinfo {pages} {275} (\bibinfo {year} {2006})}\BibitemShut {NoStop}%
\bibitem [{\citenamefont {Garc\'{\i}a}\ \emph {et~al.}(2015)\citenamefont
  {Garc\'{\i}a}, \citenamefont {Covaci},\ and\ \citenamefont
  {Rappoport}}]{Rappoport2015}%
  \BibitemOpen
  \bibfield  {author} {\bibinfo {author} {\bibfnamefont {J.~H.}\ \bibnamefont
  {Garc\'{\i}a}}, \bibinfo {author} {\bibfnamefont {L.}~\bibnamefont {Covaci}},
  \ and\ \bibinfo {author} {\bibfnamefont {T.~G.}\ \bibnamefont {Rappoport}},\
  }\href {\doibase 10.1103/PhysRevLett.114.116602} {\bibfield  {journal}
  {\bibinfo  {journal} {Phys. Rev. Lett.}\ }\textbf {\bibinfo {volume} {114}},\
  \bibinfo {pages} {116602} (\bibinfo {year} {2015})}\BibitemShut {NoStop}%
\bibitem [{\citenamefont {Carvalho}\ \emph {et~al.}(2018)\citenamefont
  {Carvalho}, \citenamefont {Garc\'{\i}a-Mart\'{\i}nez}, \citenamefont {Lado},\
  and\ \citenamefont {Fern\'andez-Rossier}}]{Carvalho2018}%
  \BibitemOpen
  \bibfield  {author} {\bibinfo {author} {\bibfnamefont {D.}~\bibnamefont
  {Carvalho}}, \bibinfo {author} {\bibfnamefont {N.~A.}\ \bibnamefont
  {Garc\'{\i}a-Mart\'{\i}nez}}, \bibinfo {author} {\bibfnamefont {J.~L.}\
  \bibnamefont {Lado}}, \ and\ \bibinfo {author} {\bibfnamefont
  {J.}~\bibnamefont {Fern\'andez-Rossier}},\ }\href {\doibase
  10.1103/PhysRevB.97.115453} {\bibfield  {journal} {\bibinfo  {journal} {Phys.
  Rev. B}\ }\textbf {\bibinfo {volume} {97}},\ \bibinfo {pages} {115453}
  (\bibinfo {year} {2018})}\BibitemShut {NoStop}%
\bibitem [{\citenamefont {Lent}\ \emph {et~al.}(1986)\citenamefont {Lent},
  \citenamefont {Bowen}, \citenamefont {Dow}, \citenamefont {Allgaier},
  \citenamefont {Sankey},\ and\ \citenamefont {Ho}}]{Lent1986}%
  \BibitemOpen
  \bibfield  {author} {\bibinfo {author} {\bibfnamefont {C.~S.}\ \bibnamefont
  {Lent}}, \bibinfo {author} {\bibfnamefont {M.~A.}\ \bibnamefont {Bowen}},
  \bibinfo {author} {\bibfnamefont {J.~D.}\ \bibnamefont {Dow}}, \bibinfo
  {author} {\bibfnamefont {R.~S.}\ \bibnamefont {Allgaier}}, \bibinfo {author}
  {\bibfnamefont {O.~F.}\ \bibnamefont {Sankey}}, \ and\ \bibinfo {author}
  {\bibfnamefont {E.~S.}\ \bibnamefont {Ho}},\ }\href {\doibase
  10.1016/0749-6036(86)90017-0} {\bibfield  {journal} {\bibinfo  {journal}
  {Superlattices and Microstructures}\ }\textbf {\bibinfo {volume} {2}},\
  \bibinfo {pages} {491} (\bibinfo {year} {1986})}\BibitemShut {NoStop}%
\bibitem [{\citenamefont {Gao}\ and\ \citenamefont {Daw}(2008)}]{Gao2008}%
  \BibitemOpen
  \bibfield  {author} {\bibinfo {author} {\bibfnamefont {X.}~\bibnamefont
  {Gao}}\ and\ \bibinfo {author} {\bibfnamefont {M.~S.}\ \bibnamefont {Daw}},\
  }\href {\doibase 10.1103/PhysRevB.77.033103} {\bibfield  {journal} {\bibinfo
  {journal} {Phys. Rev. B}\ }\textbf {\bibinfo {volume} {77}},\ \bibinfo
  {pages} {033103} (\bibinfo {year} {2008})}\BibitemShut {NoStop}%
\bibitem [{\citenamefont {{Dziawa}}\ \emph {et~al.}(2012)\citenamefont
  {{Dziawa}}, \citenamefont {{Kowalski}}, \citenamefont {{Dybko}},
  \citenamefont {{Buczko}}, \citenamefont {{Szczerbakow}}, \citenamefont
  {{Szot}}, \citenamefont {{{\L}usakowska}}, \citenamefont {{Balasubramanian}},
  \citenamefont {{Wojek}}, \citenamefont {{Berntsen}}, \citenamefont
  {{Tjernberg}},\ and\ \citenamefont {{Story}}}]{Dziawa2012}%
  \BibitemOpen
  \bibfield  {author} {\bibinfo {author} {\bibfnamefont {P.}~\bibnamefont
  {{Dziawa}}}, \bibinfo {author} {\bibfnamefont {B.~J.}\ \bibnamefont
  {{Kowalski}}}, \bibinfo {author} {\bibfnamefont {K.}~\bibnamefont {{Dybko}}},
  \bibinfo {author} {\bibfnamefont {R.}~\bibnamefont {{Buczko}}}, \bibinfo
  {author} {\bibfnamefont {A.}~\bibnamefont {{Szczerbakow}}}, \bibinfo {author}
  {\bibfnamefont {M.}~\bibnamefont {{Szot}}}, \bibinfo {author} {\bibfnamefont
  {E.}~\bibnamefont {{{\L}usakowska}}}, \bibinfo {author} {\bibfnamefont
  {T.}~\bibnamefont {{Balasubramanian}}}, \bibinfo {author} {\bibfnamefont
  {B.~M.}\ \bibnamefont {{Wojek}}}, \bibinfo {author} {\bibfnamefont {M.~H.}\
  \bibnamefont {{Berntsen}}}, \bibinfo {author} {\bibfnamefont
  {O.}~\bibnamefont {{Tjernberg}}}, \ and\ \bibinfo {author} {\bibfnamefont
  {T.}~\bibnamefont {{Story}}},\ }\href {\doibase 10.1038/nmat3449} {\bibfield
  {journal} {\bibinfo  {journal} {Nature Materials}\ }\textbf {\bibinfo
  {volume} {11}},\ \bibinfo {pages} {1023} (\bibinfo {year}
  {2012})}\BibitemShut {NoStop}%
\bibitem [{\citenamefont {Dimmock}\ \emph {et~al.}(1966)\citenamefont
  {Dimmock}, \citenamefont {Melngailis},\ and\ \citenamefont
  {Strauss}}]{Dimmock1966}%
  \BibitemOpen
  \bibfield  {author} {\bibinfo {author} {\bibfnamefont {J.~O.}\ \bibnamefont
  {Dimmock}}, \bibinfo {author} {\bibfnamefont {I.}~\bibnamefont {Melngailis}},
  \ and\ \bibinfo {author} {\bibfnamefont {A.~J.}\ \bibnamefont {Strauss}},\
  }\href {\doibase 10.1103/PhysRevLett.16.1193} {\bibfield  {journal} {\bibinfo
   {journal} {Phys. Rev. Lett.}\ }\textbf {\bibinfo {volume} {16}},\ \bibinfo
  {pages} {1193} (\bibinfo {year} {1966})}\BibitemShut {NoStop}%
\bibitem [{\citenamefont {Teo}\ \emph {et~al.}(2008)\citenamefont {Teo},
  \citenamefont {Fu},\ and\ \citenamefont {Kane}}]{Teo2008}%
  \BibitemOpen
  \bibfield  {author} {\bibinfo {author} {\bibfnamefont {J.~C.~Y.}\
  \bibnamefont {Teo}}, \bibinfo {author} {\bibfnamefont {L.}~\bibnamefont
  {Fu}}, \ and\ \bibinfo {author} {\bibfnamefont {C.~L.}\ \bibnamefont
  {Kane}},\ }\href {\doibase 10.1103/PhysRevB.78.045426} {\bibfield  {journal}
  {\bibinfo  {journal} {Physical Review B}\ }\textbf {\bibinfo {volume} {78}},\
  \bibinfo {pages} {045426} (\bibinfo {year} {2008})}\BibitemShut {NoStop}%
\bibitem [{\citenamefont {Xu}\ \emph {et~al.}(2012)\citenamefont {Xu},
  \citenamefont {Liu}, \citenamefont {Alidoust}, \citenamefont {Neupane},
  \citenamefont {Qian}, \citenamefont {Belopolski}, \citenamefont {Denlinger},
  \citenamefont {Wang}, \citenamefont {Lin}, \citenamefont {Wray},
  \citenamefont {Landolt}, \citenamefont {Slomski}, \citenamefont {Dil},
  \citenamefont {Marcinkova}, \citenamefont {Morosan}, \citenamefont {Gibson},
  \citenamefont {Sankar}, \citenamefont {Chou}, \citenamefont {Cava},
  \citenamefont {Bansil},\ and\ \citenamefont {Hasan}}]{Xu2012}%
  \BibitemOpen
  \bibfield  {author} {\bibinfo {author} {\bibfnamefont {S.-Y.}\ \bibnamefont
  {Xu}}, \bibinfo {author} {\bibfnamefont {C.}~\bibnamefont {Liu}}, \bibinfo
  {author} {\bibfnamefont {N.}~\bibnamefont {Alidoust}}, \bibinfo {author}
  {\bibfnamefont {M.}~\bibnamefont {Neupane}}, \bibinfo {author} {\bibfnamefont
  {D.}~\bibnamefont {Qian}}, \bibinfo {author} {\bibfnamefont {I.}~\bibnamefont
  {Belopolski}}, \bibinfo {author} {\bibfnamefont {J.}~\bibnamefont
  {Denlinger}}, \bibinfo {author} {\bibfnamefont {Y.}~\bibnamefont {Wang}},
  \bibinfo {author} {\bibfnamefont {H.}~\bibnamefont {Lin}}, \bibinfo {author}
  {\bibfnamefont {L.}~\bibnamefont {Wray}}, \bibinfo {author} {\bibfnamefont
  {G.}~\bibnamefont {Landolt}}, \bibinfo {author} {\bibfnamefont
  {B.}~\bibnamefont {Slomski}}, \bibinfo {author} {\bibfnamefont
  {J.}~\bibnamefont {Dil}}, \bibinfo {author} {\bibfnamefont {A.}~\bibnamefont
  {Marcinkova}}, \bibinfo {author} {\bibfnamefont {E.}~\bibnamefont {Morosan}},
  \bibinfo {author} {\bibfnamefont {Q.}~\bibnamefont {Gibson}}, \bibinfo
  {author} {\bibfnamefont {R.}~\bibnamefont {Sankar}}, \bibinfo {author}
  {\bibfnamefont {F.}~\bibnamefont {Chou}}, \bibinfo {author} {\bibfnamefont
  {R.}~\bibnamefont {Cava}}, \bibinfo {author} {\bibfnamefont {A.}~\bibnamefont
  {Bansil}}, \ and\ \bibinfo {author} {\bibfnamefont {M.}~\bibnamefont
  {Hasan}},\ }\href {\doibase 10.1038/ncomms2191} {\bibfield  {journal}
  {\bibinfo  {journal} {Nature Communications}\ }\textbf {\bibinfo {volume}
  {3}},\  (\bibinfo {year} {2012})}\BibitemShut {NoStop}%
\bibitem [{\citenamefont {Tanaka}\ \emph {et~al.}(2012)\citenamefont {Tanaka},
  \citenamefont {Ren}, \citenamefont {Sato}, \citenamefont {Nakayama},
  \citenamefont {Souma}, \citenamefont {Takahashi}, \citenamefont {Segawa},\
  and\ \citenamefont {Ando}}]{Tanaka2012}%
  \BibitemOpen
  \bibfield  {author} {\bibinfo {author} {\bibfnamefont {Y.}~\bibnamefont
  {Tanaka}}, \bibinfo {author} {\bibfnamefont {Z.}~\bibnamefont {Ren}},
  \bibinfo {author} {\bibfnamefont {T.}~\bibnamefont {Sato}}, \bibinfo {author}
  {\bibfnamefont {K.}~\bibnamefont {Nakayama}}, \bibinfo {author}
  {\bibfnamefont {S.}~\bibnamefont {Souma}}, \bibinfo {author} {\bibfnamefont
  {T.}~\bibnamefont {Takahashi}}, \bibinfo {author} {\bibfnamefont
  {K.}~\bibnamefont {Segawa}}, \ and\ \bibinfo {author} {\bibfnamefont
  {Y.}~\bibnamefont {Ando}},\ }\href {\doibase 10.1038/nphys2442} {\bibfield
  {journal} {\bibinfo  {journal} {Nature Physics}\ }\textbf {\bibinfo {volume}
  {8}},\ \bibinfo {pages} {800} (\bibinfo {year} {2012})}\BibitemShut {NoStop}%
\bibitem [{\citenamefont {Tanaka}\ \emph {et~al.}(2013)\citenamefont {Tanaka},
  \citenamefont {Sato}, \citenamefont {Nakayama}, \citenamefont {Souma},
  \citenamefont {Takahashi}, \citenamefont {Ren}, \citenamefont {Novak},
  \citenamefont {Segawa},\ and\ \citenamefont {Ando}}]{Tanaka2013}%
  \BibitemOpen
  \bibfield  {author} {\bibinfo {author} {\bibfnamefont {Y.}~\bibnamefont
  {Tanaka}}, \bibinfo {author} {\bibfnamefont {T.}~\bibnamefont {Sato}},
  \bibinfo {author} {\bibfnamefont {K.}~\bibnamefont {Nakayama}}, \bibinfo
  {author} {\bibfnamefont {S.}~\bibnamefont {Souma}}, \bibinfo {author}
  {\bibfnamefont {T.}~\bibnamefont {Takahashi}}, \bibinfo {author}
  {\bibfnamefont {Z.}~\bibnamefont {Ren}}, \bibinfo {author} {\bibfnamefont
  {M.}~\bibnamefont {Novak}}, \bibinfo {author} {\bibfnamefont
  {K.}~\bibnamefont {Segawa}}, \ and\ \bibinfo {author} {\bibfnamefont
  {Y.}~\bibnamefont {Ando}},\ }\href {\doibase 10.1103/PhysRevB.87.155105}
  {\bibfield  {journal} {\bibinfo  {journal} {Phys. Rev. B}\ }\textbf {\bibinfo
  {volume} {87}},\ \bibinfo {pages} {155105} (\bibinfo {year}
  {2013})}\BibitemShut {NoStop}%
\bibitem [{\citenamefont {Yan}\ \emph {et~al.}(2014)\citenamefont {Yan},
  \citenamefont {Liu}, \citenamefont {Zang}, \citenamefont {Wang},
  \citenamefont {Wang}, \citenamefont {Wang}, \citenamefont {Zhang},
  \citenamefont {Wang}, \citenamefont {Ma}, \citenamefont {Ji}, \citenamefont
  {He}, \citenamefont {Fu}, \citenamefont {Duan}, \citenamefont {Xue},\ and\
  \citenamefont {Chen}}]{Yan2014}%
  \BibitemOpen
  \bibfield  {author} {\bibinfo {author} {\bibfnamefont {C.}~\bibnamefont
  {Yan}}, \bibinfo {author} {\bibfnamefont {J.}~\bibnamefont {Liu}}, \bibinfo
  {author} {\bibfnamefont {Y.}~\bibnamefont {Zang}}, \bibinfo {author}
  {\bibfnamefont {J.}~\bibnamefont {Wang}}, \bibinfo {author} {\bibfnamefont
  {Z.}~\bibnamefont {Wang}}, \bibinfo {author} {\bibfnamefont {P.}~\bibnamefont
  {Wang}}, \bibinfo {author} {\bibfnamefont {Z.-D.}\ \bibnamefont {Zhang}},
  \bibinfo {author} {\bibfnamefont {L.}~\bibnamefont {Wang}}, \bibinfo {author}
  {\bibfnamefont {X.}~\bibnamefont {Ma}}, \bibinfo {author} {\bibfnamefont
  {S.}~\bibnamefont {Ji}}, \bibinfo {author} {\bibfnamefont {K.}~\bibnamefont
  {He}}, \bibinfo {author} {\bibfnamefont {L.}~\bibnamefont {Fu}}, \bibinfo
  {author} {\bibfnamefont {W.}~\bibnamefont {Duan}}, \bibinfo {author}
  {\bibfnamefont {Q.-K.}\ \bibnamefont {Xue}}, \ and\ \bibinfo {author}
  {\bibfnamefont {X.}~\bibnamefont {Chen}},\ }\href {\doibase
  10.1103/physrevlett.112.186801} {\bibfield  {journal} {\bibinfo  {journal}
  {Physical Review Letters}\ }\textbf {\bibinfo {volume} {112}},\ \bibinfo
  {pages} {186801} (\bibinfo {year} {2014})}\BibitemShut {NoStop}%
\bibitem [{\citenamefont {Zhong}\ \emph {et~al.}(2015)\citenamefont {Zhong},
  \citenamefont {He}, \citenamefont {Schneeloch}, \citenamefont {Zhang},
  \citenamefont {Liu}, \citenamefont {Pletikosi\ifmmode~\acute{c}\else
  \'{c}\fi{}}, \citenamefont {Yilmaz}, \citenamefont {Sinkovic}, \citenamefont
  {Li}, \citenamefont {Ku}, \citenamefont {Valla}, \citenamefont {Tranquada},\
  and\ \citenamefont {Gu}}]{Zhong2015}%
  \BibitemOpen
  \bibfield  {author} {\bibinfo {author} {\bibfnamefont {R.}~\bibnamefont
  {Zhong}}, \bibinfo {author} {\bibfnamefont {X.}~\bibnamefont {He}}, \bibinfo
  {author} {\bibfnamefont {J.~A.}\ \bibnamefont {Schneeloch}}, \bibinfo
  {author} {\bibfnamefont {C.}~\bibnamefont {Zhang}}, \bibinfo {author}
  {\bibfnamefont {T.}~\bibnamefont {Liu}}, \bibinfo {author} {\bibfnamefont
  {I.}~\bibnamefont {Pletikosi\ifmmode~\acute{c}\else \'{c}\fi{}}}, \bibinfo
  {author} {\bibfnamefont {T.}~\bibnamefont {Yilmaz}}, \bibinfo {author}
  {\bibfnamefont {B.}~\bibnamefont {Sinkovic}}, \bibinfo {author}
  {\bibfnamefont {Q.}~\bibnamefont {Li}}, \bibinfo {author} {\bibfnamefont
  {W.}~\bibnamefont {Ku}}, \bibinfo {author} {\bibfnamefont {T.}~\bibnamefont
  {Valla}}, \bibinfo {author} {\bibfnamefont {J.~M.}\ \bibnamefont
  {Tranquada}}, \ and\ \bibinfo {author} {\bibfnamefont {G.}~\bibnamefont
  {Gu}},\ }\href {\doibase 10.1103/PhysRevB.91.195321} {\bibfield  {journal}
  {\bibinfo  {journal} {Phys. Rev. B}\ }\textbf {\bibinfo {volume} {91}},\
  \bibinfo {pages} {195321} (\bibinfo {year} {2015})}\BibitemShut {NoStop}%
\bibitem [{\citenamefont {Avron}\ and\ \citenamefont
  {Seiler}(1985)}]{Avron1985}%
  \BibitemOpen
  \bibfield  {author} {\bibinfo {author} {\bibfnamefont {J.~E.}\ \bibnamefont
  {Avron}}\ and\ \bibinfo {author} {\bibfnamefont {R.}~\bibnamefont {Seiler}},\
  }\href {\doibase 10.1103/PhysRevLett.54.259} {\bibfield  {journal} {\bibinfo
  {journal} {Physical review letters}\ }\textbf {\bibinfo {volume} {54}},\
  \bibinfo {pages} {259} (\bibinfo {year} {1985})}\BibitemShut {NoStop}%
\bibitem [{\citenamefont {Niu}\ \emph {et~al.}(1985)\citenamefont {Niu},
  \citenamefont {Thouless},\ and\ \citenamefont {Wu}}]{Niu1985}%
  \BibitemOpen
  \bibfield  {author} {\bibinfo {author} {\bibfnamefont {Q.}~\bibnamefont
  {Niu}}, \bibinfo {author} {\bibfnamefont {D.~J.}\ \bibnamefont {Thouless}}, \
  and\ \bibinfo {author} {\bibfnamefont {Y.-S.}\ \bibnamefont {Wu}},\ }\href
  {\doibase 10.1103/PhysRevB.31.3372} {\bibfield  {journal} {\bibinfo
  {journal} {Physical Review B}\ }\textbf {\bibinfo {volume} {31}},\ \bibinfo
  {pages} {3372} (\bibinfo {year} {1985})}\BibitemShut {NoStop}%
\bibitem [{\citenamefont {Fulga}\ \emph {et~al.}(2012)\citenamefont {Fulga},
  \citenamefont {Hassler},\ and\ \citenamefont {Akhmerov}}]{Fulga2012}%
  \BibitemOpen
  \bibfield  {author} {\bibinfo {author} {\bibfnamefont {I.~C.}\ \bibnamefont
  {Fulga}}, \bibinfo {author} {\bibfnamefont {F.}~\bibnamefont {Hassler}}, \
  and\ \bibinfo {author} {\bibfnamefont {A.~R.}\ \bibnamefont {Akhmerov}},\
  }\href {\doibase 10.1103/physrevb.85.165409} {\bibfield  {journal} {\bibinfo
  {journal} {Physical Review B}\ }\textbf {\bibinfo {volume} {85}},\ \bibinfo
  {pages} {165409} (\bibinfo {year} {2012})}\BibitemShut {NoStop}%
\bibitem [{\citenamefont {George}(1973)}]{George1973}%
  \BibitemOpen
  \bibfield  {author} {\bibinfo {author} {\bibfnamefont {A.}~\bibnamefont
  {George}},\ }\href {\doibase 10.1137/0710032} {\bibfield  {journal} {\bibinfo
   {journal} {{SIAM} Journal on Numerical Analysis}\ }\textbf {\bibinfo
  {volume} {10}},\ \bibinfo {pages} {345} (\bibinfo {year} {1973})}\BibitemShut
  {NoStop}%
\bibitem [{\citenamefont {Aizenman}\ and\ \citenamefont
  {Graf}(1998)}]{Aizenman1998}%
  \BibitemOpen
  \bibfield  {author} {\bibinfo {author} {\bibfnamefont {M.}~\bibnamefont
  {Aizenman}}\ and\ \bibinfo {author} {\bibfnamefont {G.~M.}\ \bibnamefont
  {Graf}},\ }\href {http://stacks.iop.org/0305-4470/31/i=32/a=004} {\bibfield
  {journal} {\bibinfo  {journal} {Journal of Physics A: Mathematical and
  General}\ }\textbf {\bibinfo {volume} {31}},\ \bibinfo {pages} {6783}
  (\bibinfo {year} {1998})}\BibitemShut {NoStop}%
\bibitem [{\citenamefont {Prodan}\ and\ \citenamefont
  {Kohn}(2005)}]{Prodan2005}%
  \BibitemOpen
  \bibfield  {author} {\bibinfo {author} {\bibfnamefont {E.}~\bibnamefont
  {Prodan}}\ and\ \bibinfo {author} {\bibfnamefont {W.}~\bibnamefont {Kohn}},\
  }\href {\doibase 10.1073/pnas.0505436102} {\bibfield  {journal} {\bibinfo
  {journal} {Proceedings of the National Academy of Sciences}\ }\textbf
  {\bibinfo {volume} {102}},\ \bibinfo {pages} {11635} (\bibinfo {year}
  {2005})}\BibitemShut {NoStop}%
\bibitem [{\citenamefont {Song}\ and\ \citenamefont {Prodan}(2014)}]{Song2014}%
  \BibitemOpen
  \bibfield  {author} {\bibinfo {author} {\bibfnamefont {J.}~\bibnamefont
  {Song}}\ and\ \bibinfo {author} {\bibfnamefont {E.}~\bibnamefont {Prodan}},\
  }\href {\doibase 10.1103/physrevb.89.224203} {\bibfield  {journal} {\bibinfo
  {journal} {Physical Review B}\ }\textbf {\bibinfo {volume} {89}},\ \bibinfo
  {pages} {224203} (\bibinfo {year} {2014})}\BibitemShut {NoStop}%
\bibitem [{\citenamefont {Mondragon-Shem}\ \emph {et~al.}(2014)\citenamefont
  {Mondragon-Shem}, \citenamefont {Hughes}, \citenamefont {Song},\ and\
  \citenamefont {Prodan}}]{MondragonShem2014}%
  \BibitemOpen
  \bibfield  {author} {\bibinfo {author} {\bibfnamefont {I.}~\bibnamefont
  {Mondragon-Shem}}, \bibinfo {author} {\bibfnamefont {T.~L.}\ \bibnamefont
  {Hughes}}, \bibinfo {author} {\bibfnamefont {J.}~\bibnamefont {Song}}, \ and\
  \bibinfo {author} {\bibfnamefont {E.}~\bibnamefont {Prodan}},\ }\href
  {\doibase 10.1103/physrevlett.113.046802} {\bibfield  {journal} {\bibinfo
  {journal} {Physical Review Letters}\ }\textbf {\bibinfo {volume} {113}},\
  \bibinfo {pages} {046802} (\bibinfo {year} {2014})}\BibitemShut {NoStop}%
\bibitem [{\citenamefont {{Perez-Piskunow}}\ \emph {et~al.}()\citenamefont
  {{Perez-Piskunow}}, \citenamefont {Varjas}, \citenamefont {Fruchart},
  \citenamefont {Caio},\ and\ \citenamefont {Akhmerov}}]{in_preparation}%
  \BibitemOpen
  \bibfield  {author} {\bibinfo {author} {\bibfnamefont {P.}~\bibnamefont
  {{Perez-Piskunow}}}, \bibinfo {author} {\bibfnamefont {D.}~\bibnamefont
  {Varjas}}, \bibinfo {author} {\bibfnamefont {M.}~\bibnamefont {Fruchart}},
  \bibinfo {author} {\bibfnamefont {M.}~\bibnamefont {Caio}}, \ and\ \bibinfo
  {author} {\bibfnamefont {A.}~\bibnamefont {Akhmerov}},\ }\href@noop {}
  {}\bibinfo {note} {In preparation.}\BibitemShut {Stop}%
\bibitem [{\citenamefont {Hsieh}\ \emph {et~al.}(2012)\citenamefont {Hsieh},
  \citenamefont {Lin}, \citenamefont {Liu}, \citenamefont {Duan}, \citenamefont
  {Bansil},\ and\ \citenamefont {Fu}}]{Hsieh2012}%
  \BibitemOpen
  \bibfield  {author} {\bibinfo {author} {\bibfnamefont {T.~H.}\ \bibnamefont
  {Hsieh}}, \bibinfo {author} {\bibfnamefont {H.}~\bibnamefont {Lin}}, \bibinfo
  {author} {\bibfnamefont {J.}~\bibnamefont {Liu}}, \bibinfo {author}
  {\bibfnamefont {W.}~\bibnamefont {Duan}}, \bibinfo {author} {\bibfnamefont
  {A.}~\bibnamefont {Bansil}}, \ and\ \bibinfo {author} {\bibfnamefont
  {L.}~\bibnamefont {Fu}},\ }\href {\doibase 10.1038/ncomms1969} {\bibfield
  {journal} {\bibinfo  {journal} {Nature Communications}\ }\textbf {\bibinfo
  {volume} {3}},\  (\bibinfo {year} {2012})}\BibitemShut {NoStop}%
\bibitem [{Note1()}]{Note1}%
  \BibitemOpen
  \bibinfo {note} {We expect that this will not happen with a regular
  short-range correlated disorder.}\BibitemShut {Stop}%
\bibitem [{\citenamefont {Fu}\ and\ \citenamefont {Kane}(2012)}]{Fu2012}%
  \BibitemOpen
  \bibfield  {author} {\bibinfo {author} {\bibfnamefont {L.}~\bibnamefont
  {Fu}}\ and\ \bibinfo {author} {\bibfnamefont {C.~L.}\ \bibnamefont {Kane}},\
  }\href {\doibase 10.1103/physrevlett.109.246605} {\bibfield  {journal}
  {\bibinfo  {journal} {Physical Review Letters}\ }\textbf {\bibinfo {volume}
  {109}},\ \bibinfo {pages} {246605} (\bibinfo {year} {2012})}\BibitemShut
  {NoStop}%
\bibitem [{\citenamefont {Ando}\ and\ \citenamefont {Fu}(2015)}]{Ando2015}%
  \BibitemOpen
  \bibfield  {author} {\bibinfo {author} {\bibfnamefont {Y.}~\bibnamefont
  {Ando}}\ and\ \bibinfo {author} {\bibfnamefont {L.}~\bibnamefont {Fu}},\
  }\href {\doibase 10.1146/annurev-conmatphys-031214-014501} {\bibfield
  {journal} {\bibinfo  {journal} {Annual Review of Condensed Matter Physics}\
  }\textbf {\bibinfo {volume} {6}},\ \bibinfo {pages} {361} (\bibinfo {year}
  {2015})}\BibitemShut {NoStop}%
\bibitem [{\citenamefont {Okada}\ \emph {et~al.}(2013)\citenamefont {Okada},
  \citenamefont {Serbyn}, \citenamefont {Lin}, \citenamefont {Walkup},
  \citenamefont {Zhou}, \citenamefont {Dhital}, \citenamefont {Neupane},
  \citenamefont {Xu}, \citenamefont {Wang}, \citenamefont {Sankar},
  \citenamefont {Chou}, \citenamefont {Bansil}, \citenamefont {Hasan},
  \citenamefont {Wilson}, \citenamefont {Fu},\ and\ \citenamefont
  {Madhavan}}]{Okada2013}%
  \BibitemOpen
  \bibfield  {author} {\bibinfo {author} {\bibfnamefont {Y.}~\bibnamefont
  {Okada}}, \bibinfo {author} {\bibfnamefont {M.}~\bibnamefont {Serbyn}},
  \bibinfo {author} {\bibfnamefont {H.}~\bibnamefont {Lin}}, \bibinfo {author}
  {\bibfnamefont {D.}~\bibnamefont {Walkup}}, \bibinfo {author} {\bibfnamefont
  {W.}~\bibnamefont {Zhou}}, \bibinfo {author} {\bibfnamefont {C.}~\bibnamefont
  {Dhital}}, \bibinfo {author} {\bibfnamefont {M.}~\bibnamefont {Neupane}},
  \bibinfo {author} {\bibfnamefont {S.}~\bibnamefont {Xu}}, \bibinfo {author}
  {\bibfnamefont {Y.~J.}\ \bibnamefont {Wang}}, \bibinfo {author}
  {\bibfnamefont {R.}~\bibnamefont {Sankar}}, \bibinfo {author} {\bibfnamefont
  {F.}~\bibnamefont {Chou}}, \bibinfo {author} {\bibfnamefont {A.}~\bibnamefont
  {Bansil}}, \bibinfo {author} {\bibfnamefont {M.~Z.}\ \bibnamefont {Hasan}},
  \bibinfo {author} {\bibfnamefont {S.~D.}\ \bibnamefont {Wilson}}, \bibinfo
  {author} {\bibfnamefont {L.}~\bibnamefont {Fu}}, \ and\ \bibinfo {author}
  {\bibfnamefont {V.}~\bibnamefont {Madhavan}},\ }\href {\doibase
  10.1126/science.1239451} {\bibfield  {journal} {\bibinfo  {journal}
  {Science}\ }\textbf {\bibinfo {volume} {341}},\ \bibinfo {pages} {1496}
  (\bibinfo {year} {2013})}\BibitemShut {NoStop}%
\bibitem [{\citenamefont {Fang}\ \emph {et~al.}(2014)\citenamefont {Fang},
  \citenamefont {Gilbert},\ and\ \citenamefont {Bernevig}}]{Fang2014}%
  \BibitemOpen
  \bibfield  {author} {\bibinfo {author} {\bibfnamefont {C.}~\bibnamefont
  {Fang}}, \bibinfo {author} {\bibfnamefont {M.~J.}\ \bibnamefont {Gilbert}}, \
  and\ \bibinfo {author} {\bibfnamefont {B.~A.}\ \bibnamefont {Bernevig}},\
  }\href {\doibase 10.1103/PhysRevLett.112.046801} {\bibfield  {journal}
  {\bibinfo  {journal} {Phys. Rev. Lett.}\ }\textbf {\bibinfo {volume} {112}},\
  \bibinfo {pages} {046801} (\bibinfo {year} {2014})}\BibitemShut {NoStop}%
\bibitem [{\citenamefont {Mitchell}\ and\ \citenamefont
  {Wallis}(1966)}]{Mitchell1966}%
  \BibitemOpen
  \bibfield  {author} {\bibinfo {author} {\bibfnamefont {D.~L.}\ \bibnamefont
  {Mitchell}}\ and\ \bibinfo {author} {\bibfnamefont {R.~F.}\ \bibnamefont
  {Wallis}},\ }\href {\doibase 10.1103/PhysRev.151.581} {\bibfield  {journal}
  {\bibinfo  {journal} {Phys. Rev.}\ }\textbf {\bibinfo {volume} {151}},\
  \bibinfo {pages} {581} (\bibinfo {year} {1966})}\BibitemShut {NoStop}%
\bibitem [{\citenamefont {Fulga}\ \emph {et~al.}(2016)\citenamefont {Fulga},
  \citenamefont {Avraham}, \citenamefont {Beidenkopf},\ and\ \citenamefont
  {Stern}}]{Fulga2016}%
  \BibitemOpen
  \bibfield  {author} {\bibinfo {author} {\bibfnamefont {I.~C.}\ \bibnamefont
  {Fulga}}, \bibinfo {author} {\bibfnamefont {N.}~\bibnamefont {Avraham}},
  \bibinfo {author} {\bibfnamefont {H.}~\bibnamefont {Beidenkopf}}, \ and\
  \bibinfo {author} {\bibfnamefont {A.}~\bibnamefont {Stern}},\ }\href
  {\doibase 10.1103/PhysRevB.94.125405} {\bibfield  {journal} {\bibinfo
  {journal} {Phys. Rev. B}\ }\textbf {\bibinfo {volume} {94}},\ \bibinfo
  {pages} {125405} (\bibinfo {year} {2016})}\BibitemShut {NoStop}%
\bibitem [{\citenamefont {Groth}\ \emph {et~al.}(2014)\citenamefont {Groth},
  \citenamefont {Wimmer}, \citenamefont {Akhmerov},\ and\ \citenamefont
  {Waintal}}]{Groth2014}%
  \BibitemOpen
  \bibfield  {author} {\bibinfo {author} {\bibfnamefont {C.~W.}\ \bibnamefont
  {Groth}}, \bibinfo {author} {\bibfnamefont {M.}~\bibnamefont {Wimmer}},
  \bibinfo {author} {\bibfnamefont {A.~R.}\ \bibnamefont {Akhmerov}}, \ and\
  \bibinfo {author} {\bibfnamefont {X.}~\bibnamefont {Waintal}},\ }\href
  {\doibase 10.1088/1367-2630/16/6/063065} {\bibfield  {journal} {\bibinfo
  {journal} {New Journal of Physics}\ }\textbf {\bibinfo {volume} {16}},\
  \bibinfo {pages} {063065} (\bibinfo {year} {2014})}\BibitemShut {NoStop}%
\bibitem [{\citenamefont {Varjas}\ \emph {et~al.}(2019)\citenamefont {Varjas},
  \citenamefont {Fruchart}, \citenamefont {Akhmerov},\ and\ \citenamefont
  {Perez-Piskunow}}]{zenodo}%
  \BibitemOpen
  \bibfield  {author} {\bibinfo {author} {\bibfnamefont {D.}~\bibnamefont
  {Varjas}}, \bibinfo {author} {\bibfnamefont {M.}~\bibnamefont {Fruchart}},
  \bibinfo {author} {\bibfnamefont {A.~R.}\ \bibnamefont {Akhmerov}}, \ and\
  \bibinfo {author} {\bibfnamefont {P.}~\bibnamefont {Perez-Piskunow}},\ }\href
  {\doibase 10.5281/zenodo.2667604} {\bibfield  {journal} {\bibinfo  {journal}
  {zenodo.2667604}\ ,\ } (\bibinfo {year} {2019})}\BibitemShut {NoStop}%
\bibitem [{\citenamefont {Caio}\ \emph {et~al.}(2019)\citenamefont {Caio},
  \citenamefont {M{\"o}ller}, \citenamefont {Cooper},\ and\ \citenamefont
  {Bhaseen}}]{Caio2019}%
  \BibitemOpen
  \bibfield  {author} {\bibinfo {author} {\bibfnamefont {M.~D.}\ \bibnamefont
  {Caio}}, \bibinfo {author} {\bibfnamefont {G.}~\bibnamefont {M{\"o}ller}},
  \bibinfo {author} {\bibfnamefont {N.~R.}\ \bibnamefont {Cooper}}, \ and\
  \bibinfo {author} {\bibfnamefont {M.~J.}\ \bibnamefont {Bhaseen}},\ }\href
  {\doibase 10.1038/s41567-018-0390-7} {\bibfield  {journal} {\bibinfo
  {journal} {Nature Physics}\ ,\ } (\bibinfo {year} {2019})}\BibitemShut
  {NoStop}%
\bibitem [{\citenamefont {Phuphachong}\ \emph {et~al.}(2017)\citenamefont
  {Phuphachong}, \citenamefont {Assaf}, \citenamefont {Volobuev}, \citenamefont
  {Bauer}, \citenamefont {Springholz}, \citenamefont {{De Vaulchier}},\ and\
  \citenamefont {Guldner}}]{Phuphachong2017}%
  \BibitemOpen
  \bibfield  {author} {\bibinfo {author} {\bibfnamefont {T.}~\bibnamefont
  {Phuphachong}}, \bibinfo {author} {\bibfnamefont {B.~A.}\ \bibnamefont
  {Assaf}}, \bibinfo {author} {\bibfnamefont {V.~V.}\ \bibnamefont {Volobuev}},
  \bibinfo {author} {\bibfnamefont {G.}~\bibnamefont {Bauer}}, \bibinfo
  {author} {\bibfnamefont {G.}~\bibnamefont {Springholz}}, \bibinfo {author}
  {\bibfnamefont {L.~A.}\ \bibnamefont {{De Vaulchier}}}, \ and\ \bibinfo
  {author} {\bibfnamefont {Y.}~\bibnamefont {Guldner}},\ }\href {\doibase
  10.1088/1742-6596/864/1/012038} {\bibfield  {journal} {\bibinfo  {journal}
  {Journal of Physics: Conference Series}\ }\textbf {\bibinfo {volume} {864}}
  (\bibinfo {year} {2017}),\ 10.1088/1742-6596/864/1/012038}\BibitemShut
  {NoStop}%
\bibitem [{Note2()}]{Note2}%
  \BibitemOpen
  \bibinfo {note} {In the case of Ref.~\cite {Yan2014}, the ARPES data for
  $x=0.3$ is not conclusive, but it is interpreted with the help of VCA theory
  as a trivial phase. We claim the that for $x=0.3$, the experimental data and
  the theoretical results presented in~\cite {Yan2014} are not sufficient to
  ensure that $x=0.3$ lies on the trivial phase.}\BibitemShut {Stop}%
\bibitem [{\citenamefont {Stützer}\ \emph {et~al.}(2018)\citenamefont
  {Stützer}, \citenamefont {Plotnik}, \citenamefont {Lumer}, \citenamefont
  {Titum}, \citenamefont {H.~Lindner}, \citenamefont {Segev}, \citenamefont
  {C.~Rechtsman},\ and\ \citenamefont {Szameit}}]{Stutzer2018}%
  \BibitemOpen
  \bibfield  {author} {\bibinfo {author} {\bibfnamefont {S.}~\bibnamefont
  {Stützer}}, \bibinfo {author} {\bibfnamefont {Y.}~\bibnamefont {Plotnik}},
  \bibinfo {author} {\bibfnamefont {Y.}~\bibnamefont {Lumer}}, \bibinfo
  {author} {\bibfnamefont {P.}~\bibnamefont {Titum}}, \bibinfo {author}
  {\bibfnamefont {N.}~\bibnamefont {H.~Lindner}}, \bibinfo {author}
  {\bibfnamefont {M.}~\bibnamefont {Segev}}, \bibinfo {author} {\bibfnamefont
  {M.}~\bibnamefont {C.~Rechtsman}}, \ and\ \bibinfo {author} {\bibfnamefont
  {A.}~\bibnamefont {Szameit}},\ }\href {\doibase 10.1038/s41586-018-0418-2}
  {\bibfield  {journal} {\bibinfo  {journal} {Nature}\ }\textbf {\bibinfo
  {volume} {560}},\  (\bibinfo {year} {2018})}\BibitemShut {NoStop}%
\bibitem [{\citenamefont {Mitchell}\ \emph {et~al.}(2018)\citenamefont
  {Mitchell}, \citenamefont {Nash}, \citenamefont {Hexner}, \citenamefont
  {Turner},\ and\ \citenamefont {Irvine}}]{Mitchell2018}%
  \BibitemOpen
  \bibfield  {author} {\bibinfo {author} {\bibfnamefont {N.~P.}\ \bibnamefont
  {Mitchell}}, \bibinfo {author} {\bibfnamefont {L.~M.}\ \bibnamefont {Nash}},
  \bibinfo {author} {\bibfnamefont {D.}~\bibnamefont {Hexner}}, \bibinfo
  {author} {\bibfnamefont {A.~M.}\ \bibnamefont {Turner}}, \ and\ \bibinfo
  {author} {\bibfnamefont {W.~T.~M.}\ \bibnamefont {Irvine}},\ }\href {\doibase
  10.1038/s41567-017-0024-5} {\bibfield  {journal} {\bibinfo  {journal} {Nature
  Physics}\ }\textbf {\bibinfo {volume} {14}},\ \bibinfo {pages} {380}
  (\bibinfo {year} {2018})}\BibitemShut {NoStop}%
\bibitem [{\citenamefont {Daraio}\ \emph {et~al.}(2018)\citenamefont {Daraio},
  \citenamefont {Cha},\ and\ \citenamefont {Woo~Kim}}]{Daraio2018}%
  \BibitemOpen
  \bibfield  {author} {\bibinfo {author} {\bibfnamefont {C.}~\bibnamefont
  {Daraio}}, \bibinfo {author} {\bibfnamefont {J.}~\bibnamefont {Cha}}, \ and\
  \bibinfo {author} {\bibfnamefont {K.}~\bibnamefont {Woo~Kim}},\ }\href
  {\doibase 10.1038/s41586-018-0764-0} {\bibfield  {journal} {\bibinfo
  {journal} {Nature}\ }\textbf {\bibinfo {volume} {564}},\  (\bibinfo {year}
  {2018})}\BibitemShut {NoStop}%
\bibitem [{\citenamefont {Zhou}\ \emph {et~al.}(2018)\citenamefont {Zhou},
  \citenamefont {Zhang},\ and\ \citenamefont {Mao}}]{Zhou2018}%
  \BibitemOpen
  \bibfield  {author} {\bibinfo {author} {\bibfnamefont {D.}~\bibnamefont
  {Zhou}}, \bibinfo {author} {\bibfnamefont {L.}~\bibnamefont {Zhang}}, \ and\
  \bibinfo {author} {\bibfnamefont {X.}~\bibnamefont {Mao}},\ }\href {\doibase
  10.1103/physrevlett.120.068003} {\bibfield  {journal} {\bibinfo  {journal}
  {Physical Review Letters}\ }\textbf {\bibinfo {volume} {120}},\ \bibinfo
  {pages} {068003} (\bibinfo {year} {2018})}\BibitemShut {NoStop}%
\bibitem [{\citenamefont {Bianco}\ and\ \citenamefont
  {Resta}(2013)}]{Bianco2013}%
  \BibitemOpen
  \bibfield  {author} {\bibinfo {author} {\bibfnamefont {R.}~\bibnamefont
  {Bianco}}\ and\ \bibinfo {author} {\bibfnamefont {R.}~\bibnamefont {Resta}},\
  }\href {\doibase 10.1103/PhysRevLett.110.087202} {\bibfield  {journal}
  {\bibinfo  {journal} {Phys. Rev. Lett.}\ }\textbf {\bibinfo {volume} {110}},\
  \bibinfo {pages} {087202} (\bibinfo {year} {2013})}\BibitemShut {NoStop}%
\bibitem [{\citenamefont {Benalcazar}\ \emph {et~al.}(2017)\citenamefont
  {Benalcazar}, \citenamefont {Bernevig},\ and\ \citenamefont
  {Hughes}}]{Benalcazar2017}%
  \BibitemOpen
  \bibfield  {author} {\bibinfo {author} {\bibfnamefont {W.~A.}\ \bibnamefont
  {Benalcazar}}, \bibinfo {author} {\bibfnamefont {B.~A.}\ \bibnamefont
  {Bernevig}}, \ and\ \bibinfo {author} {\bibfnamefont {T.~L.}\ \bibnamefont
  {Hughes}},\ }\href {\doibase 10.1126/science.aah6442} {\bibfield  {journal}
  {\bibinfo  {journal} {Science}\ }\textbf {\bibinfo {volume} {357}},\ \bibinfo
  {pages} {61} (\bibinfo {year} {2017})}\BibitemShut {NoStop}%
\bibitem [{\citenamefont {Slater}\ and\ \citenamefont
  {Koster}(1954)}]{Slater1954}%
  \BibitemOpen
  \bibfield  {author} {\bibinfo {author} {\bibfnamefont {J.~C.}\ \bibnamefont
  {Slater}}\ and\ \bibinfo {author} {\bibfnamefont {G.~F.}\ \bibnamefont
  {Koster}},\ }\href {\doibase 10.1103/PhysRev.94.1498} {\bibfield  {journal}
  {\bibinfo  {journal} {Phys. Rev.}\ }\textbf {\bibinfo {volume} {94}},\
  \bibinfo {pages} {1498} (\bibinfo {year} {1954})}\BibitemShut {NoStop}%
\bibitem [{\citenamefont {Schindler}\ \emph {et~al.}(2018)\citenamefont
  {Schindler}, \citenamefont {Cook}, \citenamefont {Vergniory}, \citenamefont
  {Wang}, \citenamefont {Parkin}, \citenamefont {Bernevig},\ and\ \citenamefont
  {Neupert}}]{Schindler2018}%
  \BibitemOpen
  \bibfield  {author} {\bibinfo {author} {\bibfnamefont {F.}~\bibnamefont
  {Schindler}}, \bibinfo {author} {\bibfnamefont {A.~M.}\ \bibnamefont {Cook}},
  \bibinfo {author} {\bibfnamefont {M.~G.}\ \bibnamefont {Vergniory}}, \bibinfo
  {author} {\bibfnamefont {Z.}~\bibnamefont {Wang}}, \bibinfo {author}
  {\bibfnamefont {S.~S.~P.}\ \bibnamefont {Parkin}}, \bibinfo {author}
  {\bibfnamefont {B.~A.}\ \bibnamefont {Bernevig}}, \ and\ \bibinfo {author}
  {\bibfnamefont {T.}~\bibnamefont {Neupert}},\ }\href {\doibase
  10.1126/sciadv.aat0346} {\bibfield  {journal} {\bibinfo  {journal} {Science
  Advances}\ }\textbf {\bibinfo {volume} {4}},\  (\bibinfo {year}
  {2018})}\BibitemShut {NoStop}%
\bibitem [{\citenamefont {Sessi}\ \emph {et~al.}(2016)\citenamefont {Sessi},
  \citenamefont {Di~Sante}, \citenamefont {Szczerbakow}, \citenamefont {Glott},
  \citenamefont {Wilfert}, \citenamefont {Schmidt}, \citenamefont {Bathon},
  \citenamefont {Dziawa}, \citenamefont {Greiter}, \citenamefont {Neupert},
  \citenamefont {Sangiovanni}, \citenamefont {Story}, \citenamefont {Thomale},\
  and\ \citenamefont {Bode}}]{Sessi2016}%
  \BibitemOpen
  \bibfield  {author} {\bibinfo {author} {\bibfnamefont {P.}~\bibnamefont
  {Sessi}}, \bibinfo {author} {\bibfnamefont {D.}~\bibnamefont {Di~Sante}},
  \bibinfo {author} {\bibfnamefont {A.}~\bibnamefont {Szczerbakow}}, \bibinfo
  {author} {\bibfnamefont {F.}~\bibnamefont {Glott}}, \bibinfo {author}
  {\bibfnamefont {S.}~\bibnamefont {Wilfert}}, \bibinfo {author} {\bibfnamefont
  {H.}~\bibnamefont {Schmidt}}, \bibinfo {author} {\bibfnamefont
  {T.}~\bibnamefont {Bathon}}, \bibinfo {author} {\bibfnamefont
  {P.}~\bibnamefont {Dziawa}}, \bibinfo {author} {\bibfnamefont
  {M.}~\bibnamefont {Greiter}}, \bibinfo {author} {\bibfnamefont
  {T.}~\bibnamefont {Neupert}}, \bibinfo {author} {\bibfnamefont
  {G.}~\bibnamefont {Sangiovanni}}, \bibinfo {author} {\bibfnamefont
  {T.}~\bibnamefont {Story}}, \bibinfo {author} {\bibfnamefont
  {R.}~\bibnamefont {Thomale}}, \ and\ \bibinfo {author} {\bibfnamefont
  {M.}~\bibnamefont {Bode}},\ }\href {\doibase 10.1126/science.aah6233}
  {\bibfield  {journal} {\bibinfo  {journal} {Science}\ }\textbf {\bibinfo
  {volume} {354}},\ \bibinfo {pages} {1269} (\bibinfo {year}
  {2016})}\BibitemShut {NoStop}%
\end{thebibliography}%

\appendix

\section{Spectral projector operator expanded with the kernel polynomial method}
\label{app:KPM}
\begin{figure}
    \includegraphics[width=0.9\columnwidth]{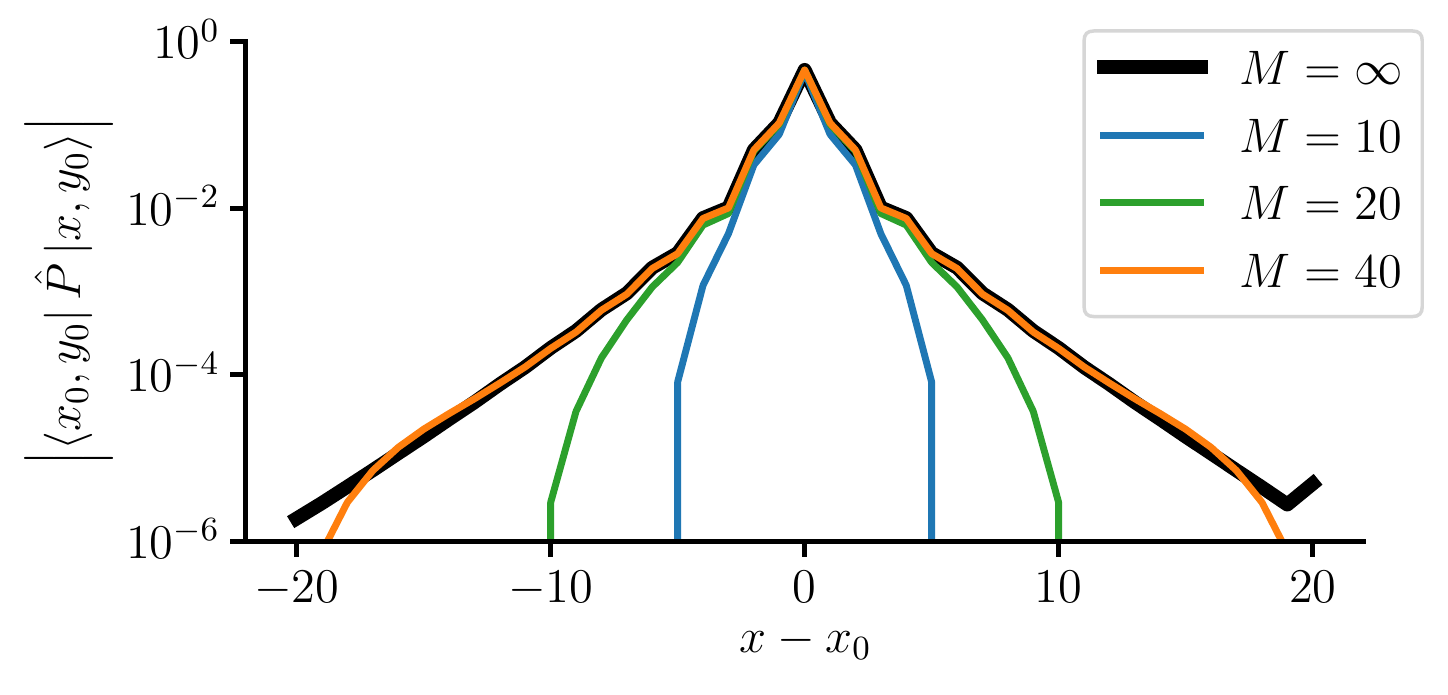}
    \caption{Convergence of the projector operator, computed via a KPM expansion with $M$ moments, applied to a vector $\lvert x_0, y_0\rangle$ located at the center of a Chern insulator.
    The KPM expansion converges to the exact projector operator for $M=\infty$, and finite values of $M$ yield finite range approximations of the projector.
    The error of the approximation scales with $\exp(-\xi/M)$~\cite{Weisse2006}.}
    \label{fig:projector}
\end{figure}

The band projector is expanded with the kernel polynomial method (KPM) \cite{Weisse2006}
and it is a finite range approximation, where the range depends linearly on the number of moments used in the expansion. Each moment of the expansion of the projector operator applied to a vector is obtained recursively, by applying the Hamiltonian to expanded vector of the previous iteration. The recursive algorithm effectively \emph{spreads} a local vector to the neighboring sites via the hopping terms of the Hamiltonian, as depicted in Fig.~\ref{fig:projector}.

Concurrently, each order of the expansion increases the precision in energy of the approximation, therefore, setting an energy resolution for the expansion.
To apply KPM, operators are rescaled such that the spectrum is in the $[-1, 1]$ range.
As a consequence, the order of the expansion required to resolve the mobility gap is proportional to $W/\Delta$, the full bandwidth divided by the mobility gap.
This can be a large ratio even if $\xi$ is small, for example in an atomic insulator with a large range of on-site energies and vanishing hoppings.
Typically $W/\Delta$ and $\xi$ increase together near a phase transition where the gap closes, but the bandwidth changes slowly, and does not affect the scaling of the computational cost with $\xi$.

The kernel polynomial method provides a stable and efficient method to expand the action of
any function of an operator $\hat{f}$ that depends on the Hamiltonian $H$ and a set of parameters $\lambda$, on a vector $\ket{v}$~\cite{Weisse2006,Rappoport2015}. The expansion up to order $M$ is
\begin{align*}
\hat{f}(\lambda, H) \ket{v} &= \sum_{m=0}^M \mu_m(\lambda) T_m (H) \ket{v}\\
&= \sum_{m=0}^M \mu_m(\lambda) \ket{v_m},
\end{align*}

The coefficients, called \emph{moments} in the context of KPM expansions, are defined as
\begin{align*}
    \mu_m(\lambda) = \frac{2}{\pi}\frac{1}{1+\delta_{m,0}} \int_{-1}^{1}
    \frac{\hat{f}(\lambda,E)T_m(E)}{\sqrt{1-E^2}}\mathrm{d}E,
\end{align*}
and the vectors $\ket{v_m}$ satisfy the recursion relation
\begin{align*}
\ket{v_0} &= \ket{v} \\
\ket{v_1} &= H \ket{v_0} \\
\ket{v_{m+1}} &= 2 H \ket{v_m} - \ket{v_{m-1}}.
\end{align*}

We approximate the projector operator defined as the step function
\begin{align*}
    \hat{P}(\varepsilon, H) &= \theta(\varepsilon - H)\\
    \hat{P}(\varepsilon, H) &= \sum_{m=0}^M \mu_m(\varepsilon)T_m(H),
\end{align*}
and in this case, the coefficients take the form
\begin{align*}
    \mu_m(\varepsilon) = \begin{cases}
                             1 - \frac{1}{\arccos{(\varepsilon)}} & m=0\\
                             \frac{-2\sin{\left(m\arccos{(\varepsilon)}\right)}}{m\pi} & m\ne 0
                         \end{cases}
\end{align*}

Equipped with the KPM expanded projector, we proceed to evaluate matrix elements of topological markers.
These are finite polynomials of $\hat{P}$ and other sparse operators such as position and mirror.
The matrix elements are evaluated by successive application of these operators to the states.
The resulting memory cost scales linearly with the system size (number of degrees of freedom), by cumulatively summing up the expanded vectors for fixed $E_F$, only a small number of sparse matrices and dense vectors are stored at any given time.
Most of the time cost comes from sparse matrix-vector multiplications, linear in the system size.
The number of operations is proportional to the number of moments $M$.

\section{Scaling of stochastic trace}
\label{app:stochastic_trace}

We optimize the calculation further by utilizing the stochastic trace approximation to evaluate the trace.
We take $R$ independent random phase vectors $\ket{r_i}$ that are only nonzero inside the region $S$, $\left\langle\bm{x}, l | r_i \right\rangle = \delta_{\bm{x}\in S}\exp(\ii \phi_{\bm{x}, l, i})$ with $\phi_{\bm{x}, l, i} \in [0, 2\pi]$ independent random phases for all sites and orbitals.
The trace of an operator $\hat{\mathcal{O}}$ equals the expectation value
\begin{equation}
    \Tr \hat{\mathcal{O}} = \mathbbm{E}\left( \frac{1}{R}\sum_{i=1}^R \bra{r_i} \hat{\mathcal{O}} \ket{r_i} \right) = \mathbbm{E}\left(\Tr_{\rm st} \hat{\mathcal{O}}\right),
\end{equation}
where $\mathbbm{E}$ denotes the expectation value over random vector realizations and we introduced the notation $\Tr_{\rm st} \hat{\mathcal{O}}$ for the random variable giving the stochastic trace of operator $\hat{\mathcal{O}}$.
The above equality is proved by using that the random phases are independent, hence $\mathbbm{E}\left(e^{\ii (\phi_{\bm{x}, l} - \phi_{\bm{x'}, l'})} \right) = \delta_{\bm{x}, \bm{x}'} \delta_{l,l'}$ and only the diagonal entries contribute to the expectation value.

The standard deviation of stochastic trace of an operator scales with the total square magnitude of the off-diagonal entries which enter in the expectation value with random phases~\cite{Weisse2006}:
\begin{equation}
\sigma(\Tr_{\rm st} \hat{\mathcal{O}}) = \sqrt{\frac{1}{R}\sum_{i\neq j} |\hat{\mathcal{O}}_{ij}|^2},
\end{equation}
where $\sigma(X) = \sqrt{\mathbbm{E}(|X|^2) - |\mathbbm{E}(X)|^2}$ is the standard deviation and we used that $\sigma(e^{\ii \phi_{\bm{x}, l}}) = 1$.
We also use that the standard deviation of the sum of independent random variables obeys $\sigma\left(\sum_i X_i \right) = \sqrt{\sum_i \sigma\left(X_i\right)^2}$.

We are concerned with the stochastic trace of topological markers, such as the Chern and mirror Chern operators.
In order to draw conclusions, we need to know the scaling of the off-diagonal matrix elements with respect to the relevant length scales in the problem.
There are three length scales, the lattice constant $a$, the localization length $\xi$ and the system size $L$ (this we take to be the linear size of the subsystem where we take the partial trace, the overall system size is a constant factor larger).
We use units of $a$ to measure the other two distances, and, as explained in the main text, we are interested in systems whose size is proportional to the localization length, so we will set $L = c \xi$ in the end.
We introduce $a$ as the lattice constant here for clarity, but it is an arbitrary reference length scale we can define in fully disordered (e.g. amorphous) systems as well, for example as the typical spacing of sites. As it cancels from the final result, this argument does not rely on the assumption of an underlying regular lattice.

\begin{figure}
    \includegraphics[width=0.95\linewidth]{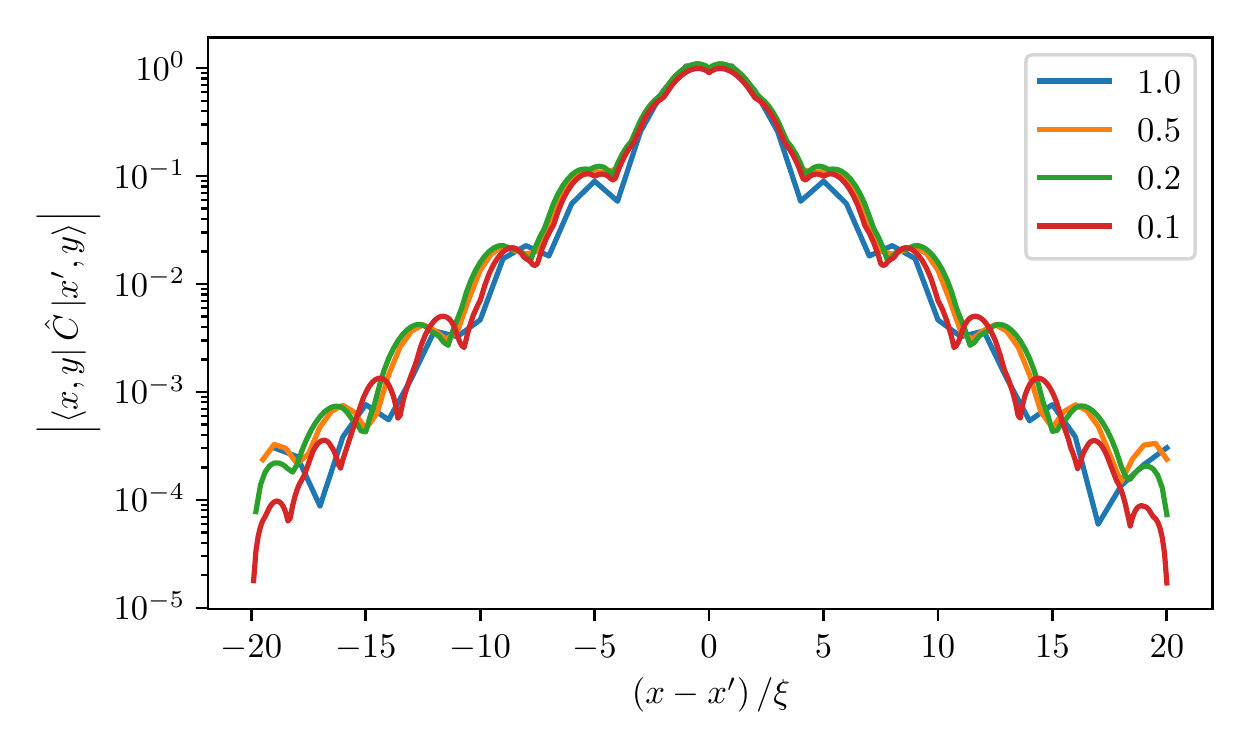}
    \caption{Off-diagonal matrix elements of the Chern operator in 2D as a function of real space distance in the units of the localization length.
				Here we use a simple continuum model of a Chern insulator, discretized on a square lattice with various lattice constants $a$ and fixed $\xi$.
				The collapse of the curves verifies the scaling form of the matrix elements that we use.
    \label{fig:chern_scaling}}
\end{figure}

We start with the Chern operator in 2D
\begin{equation}
\hat{C} = \frac{2 \pi \ii}{a^2} \left[\hat{P}\hat{x}\hat{P}, \hat{P}\hat{y}\hat{P} \right],
\end{equation}
where the $a^{-2}$ prefactor is included to measure all distances in units of $a$, this way the sum of the diagonal entries of $\hat{C}$ on a site coincides with the Chern number in a clean system.
We numerically verify (see Fig.~\ref{fig:chern_scaling}) that the off-diagonal matrix elements scale as
\begin{equation}
\bra{x, y} \hat{C} \ket{x', y'} = f\left(\frac{x-x'}{\xi}, \frac{y-y'}{\xi} \right),
\label{eq:C_scaling}
\end{equation}
where $f$ is a dimensionless function and we suppressed the dependence on the internal degrees of freedom.
$f$ is a quickly decaying function for $(x-x')/\xi \gg 1$ in insulating systems, as matrix elements of $\hat{P}$ also decay at the length scale of $\xi$.

The Chern marker averaged over a square region $S$ of size $L$ around the origin is given by
\begin{equation}
C  = \Tr_{\rm st}\left[\left( \frac{a}{L} \right)^2 \hat{C}\right],
\end{equation}
where we still measure length in units of $a$.
Substituting \eqref{eq:C_scaling} we find for the standard deviation of $C$
\begin{eqnarray}
\sigma\left(C\right) &=& \frac{1}{\sqrt{R}} \sqrt{\sum_{\bm{r}\neq\bm{r}' \in S} \left| \left(\frac{a}{L}\right)^2 f\left(\frac{\bm{r}-\bm{r}' }{\xi} \right) \right|^2} \nonumber\\
&=&  \frac{1}{\sqrt{R}} \sqrt{ \left(\frac{a}{L}\right)^4 \frac{1}{a^4} \int_S d^2 \bm{r} d^2 \bm{r}'\left| f\left(\frac{\bm{r}-\bm{r}' }{\xi} \right) \right|^2} \nonumber\\
&=&  \frac{1}{\sqrt{R}} \sqrt{ \frac{1}{L^4} \xi^4 \int_{-c/2}^{c/2} d^2 \tilde{\bm{r}}\; d^2 \tilde{\bm{r}}'\left| f\left(\tilde{\bm{r}}-\tilde{\bm{r}}' \right) \right|^2} \nonumber\\
&=& \frac{1}{c^{2}\sqrt{R}} \sqrt{\int_{-c/2}^{c/2} d^2 \tilde{\bm{r}}\; d^2 \tilde{\bm{r}}'\left| f\left(\tilde{\bm{r}}-\tilde{\bm{r}}' \right) \right|^2 }.
\end{eqnarray}
In the second line we took the limit of $\xi \gg a$, so we can replace sums over sites with integrals as $\sum_{x \in [-L/2, L/2]} = 1/a \int_{-L/2}^{L/2} dx$.
In the third line we changed the integration variables to $\tilde{\bm{r}} = \bm{r} / \xi$.
The result is only dependent on the ratio of the localization length and the system size $c = L/\xi$ and the number of random vectors $R$, but not $\xi$.
As the integral is proportional to the integration area, $c^2$ for $c \gg 1$, the overall scaling of the error with $c$ is $1/\sqrt{R c^2}$.

We find a similar scaling for other topological markers, such as the 3D winding number in chiral classes~\cite{Song2014, MondragonShem2014}.
In general, the standard deviation of the stochastic trace evaluation of the invariant depends only on the ratio of the system size and the localization length:
\begin{equation}
\sigma(\hat{\nu}) \propto \sqrt{\frac{1}{R} \left(\frac{\xi}{L}\right)^d}.
\end{equation}

\begin{figure}
    \includegraphics[width=0.95\linewidth]{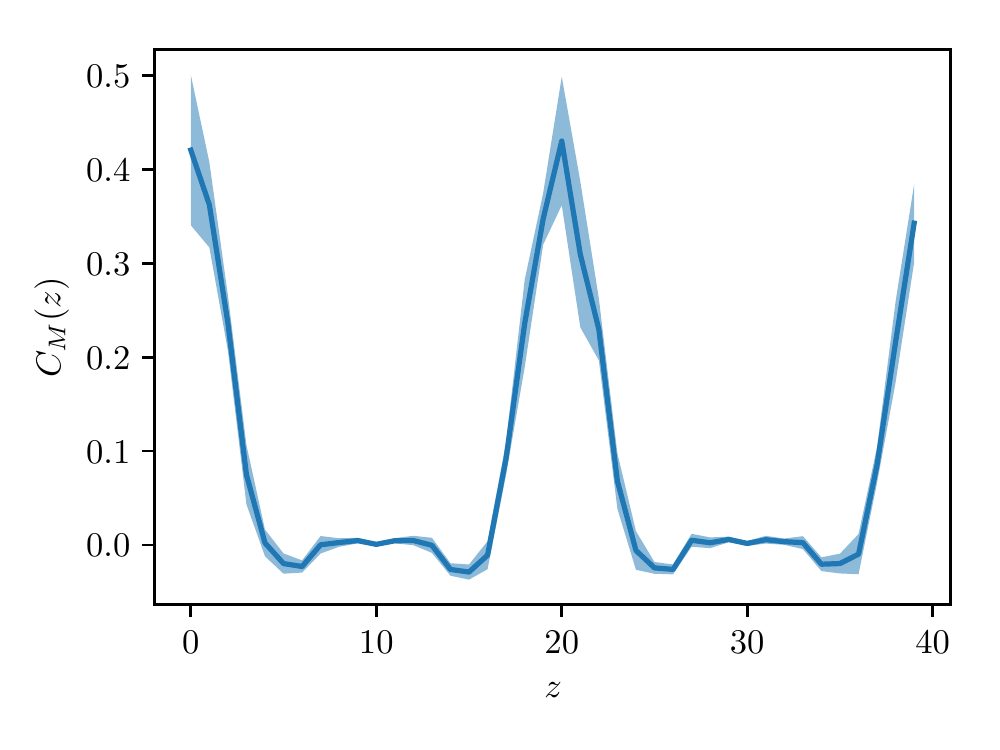}
    \caption{Contributions per layer to the mirror Chern number. The shade represents the standard deviation of the stochastic trace per layer, calculated with $10$ random vectors per layer.\label{fig:mirror_chern_layer}}
\end{figure}

For the mirror Chern operator
\begin{equation}
\hat{C}_M = \frac{\pi}{a^2} \hat{M}_z \left[\hat{P}\hat{x}\hat{P}, \hat{P}\hat{y}\hat{P} \right]
\end{equation}
the scaling is
\begin{equation}
\bra{x, y, z} \hat{C}_M \ket{x', y', z'} = \frac{a}{\xi_z} f\left(\frac{x-x'}{\xi}, \frac{y-y'}{\xi}, \frac{z+z'}{\xi_z} \right),
\end{equation}
where $\xi_z$ is the localization length in the $z$ direction.
This form is justified by the fact, that the contributions to the mirror Chern number are centered on the invariant planes, but are spread out on layers in a thickness proportional to $\xi_z$, see Fig.~\ref{fig:mirror_chern_layer}.
The total for all layers is, however, constant, hence the $a / \xi_z$ prefactor.
This is the key difference compared to the Chern number, the mirror Chern number is effectively a 2D invariant that we evaluate on a thick slab.
In a clean system every plane parallel to a mirror plane is also a mirror plane, hence the matrix element can only depend on $z+z'$.
The mirror Chern marker averaged over a square region of size $L$ is given by
\begin{equation}
C_M = \Tr_{\rm st}\left[\left( \frac{a}{L} \right)^2 \hat{C}_M\right],
\end{equation}
and we find using a similar derivation for the standard deviation (setting $\xi = \xi_z = L/c$)
\begin{equation}
\sigma\left(C_M\right) = \frac{1}{c^{2}\sqrt{R}} \sqrt{\int_{-c/2}^{c/2} d^3 \tilde{\bm{r}} \; d^3 \tilde{\bm{r}}' \left| f(\tilde{\bm{r}}, \tilde{\bm{r}}') \right|^2 },
\end{equation}
which is only dependent on the ratio of the localization length and the system size $c$ and the number of random vectors $R$.

In the numerical calculations we split the stochastic trace in two halves, using two sets of random vectors, each localized in one half of the system separated by mirror planes.
This eliminates most of the large off-diagonal entries with $z = -z'$, resulting in a constant factor reduction in the error.
Splitting the stochastic trace into more regions (e.g. separate for each layer parallel to the mirror plane) results in further reduction in the error, at the cost of increased computational effort.
The overall scaling of the computational time with $\xi$ for a fixed standard deviation is the same up to a constant factor for all of these schemes, $\xi^{d+1}$.

\section{Geometry used in the numerics}
\label{app:geometry}

\begin{figure}
    \includegraphics[width=0.8\columnwidth]{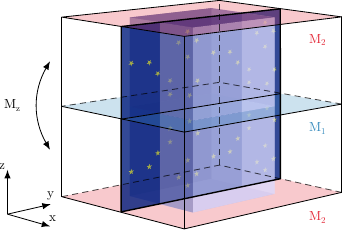}
    \caption{\label{fig:geometry_mirror}
    Geometry used for calculating the mirror Chern number.
    The upper and lower halves of the slab have mirror image disorder configurations with mirror plane $M_1$.
    Because of the PBC in the $z$ direction, there is a second mirror plane $M_2$.
    The central pane with stars is a schematic representation of the disorder, repeated in the $x$ and $y$ directions, and mirrored in $z$.
    The box in the center of the sample shows the averaging region where the partial trace is evaluated, containing one period of the disorder configuration in the $x$ and $y$ directions.
    }
\end{figure}

As described in Section~\ref{sec:mirror_chern} of the main text, we build a tight-binding model with PBC using translation vectors $W [1, 1, 0]$, $L_{1\overline{1}0} [1, -1, 0]$ and $L_z [0, 0, 1]$.
This geometry preserves the reflection symmetry with $[1, 1, 0]$ normal and contains $W\times L_{1\overline{1}0} \times L_z$ unit cells with 36 degrees of freedom each.
The averaging region of the stochastic trace extends the full width of the system in the $[1, 1, 0]$ direction and contains half of the linear size in the perpendicular directions, as depicted by the inner box in Fig.~\ref{fig:geometry_mirror}. Imposing PBC in all directions eliminates gapless surface states, and the (mobility) gap guarantees that the Fermi projector is short-ranged.

However, imposing PBC in the direction normal to the mirror planes results in two mirror invariant planes.
As argued in the main text, in a sample with open boundary conditions in the other directions, this results in a doubling of the interface modes, around the edges of the mirror invariant planes.
On the other hand, $C_M$ counts helical modes, while the 3D mirror Chern number is given by the number of chiral modes, this factor of $1/2$ cancels the previous factor of $2$.
We conclude that Eq.~\eqref{eqn:mirror_chern} is applicable to this 3D $M_z$ symmetric geometry with PBC in $z$.

\section{18-orbital tight-binding model}
\label{app:18orb}
\co{specify 18-orbital model}
We use the 18-orbital model of \ce{SnTe} and \ce{PbTe} derived in~\cite{Lent1986}.
The cubic rock-salt structure has two sublattices A and C (referring to the anion and cation nature of the atoms occupying them), the first occupied by \ce{Te} and the second by \ce{Sn} or \ce{Pb} atoms.
Each site hosts spinful $s$, $p$ and $d$ orbitals, 18 degrees of freedom in total, with annihilation operators $\bc_l$ ($l = 0, \; 1, \; 2$ for $s$, $p$, $d$ orbitals respectively), which is a vector of length $2$, $6$ or $10$ depending on the value of $l$.

The hopping terms are expressed as two-center integrals $H_{ll'm\bm{d}}$ in the \emph{linear combination of atomic orbitals} (LCAO) method\cite{Slater1954}, where $l$ and $l'$ is the total angular momentum of the orbitals connected on the two sites and $m$ is the angular momentum of the bonding along the bonding axis $\bm{d}$ ($m = 0, \; 1, \; 2$ for $\sigma$, $\pi$, $\delta$ bonding respectively).
The matrices $H_{ll'm\bm{d}}$ are $2(2l+1) \times 2(2l'+1)$ and are proportional to the identity in spin space.
The onsite terms contain different onsite energies for the various orbitals $E_l$ and ${\bm{L} \cdot \bm{S}}$ SOC terms with strength $\lambda_l$.
The tight-binding Hamiltonian reads:
\begin{multline}
\label{eq:18band}
H=  \sum_{l, \br} E_{l\br} \bc^{\dag}_{l\br} \cdot \bc_{l\br} + \sum_{l, \br} \lambda_{l \br} \bc^{\dag}_{l \br} \left(\bm{L}_l \cdot \bm{S}\right) \bc_{l \br} \\
+\sum_{l, l', m, \left\langle\br, \br' \right\rangle} V_{l, l', m, \br, \br'}  \bc^{\dag}_{l' \br'} H_{l'lm(\br' - \br)} \bc_{l \br}.
\end{multline}

The first term is the onsite energy, and it is the main source of disorder in our simulation.
For sites on the $C$ sublattice the type of the site (\ce{Sn} or \ce{Pb}) is chosen randomly with probability $1-x$ and $x$.
The value of $E_{l\br}$ is assigned accordingly to be $E_{lc}^{\text{SnTe}}$ and $E_{lc}^{\text{PbTe}}$ respectively.
The superscripts \ce{SnTe} and \ce{PbTe} refer to the two sets of parameters for the two pure materials.
If $\br\in A$, we use a weighted average $E_{l\br} = \left[n E_{la}^{\text{SnTe}} + (6 - n) E_{la}^{\text{PbTe}} \right]/6$ where $n$ is the number of nearest neighbor sites occupied by \ce{Sn} atoms.
The second term is the ${\bm{L} \cdot \bm{S}}$ spin-orbit coupling, $\bm{L}_l$ is the vector of angular momentum-$l$ operators ($0$ for $l=0$).
The values of $\lambda_{l\br}$ are assigned in the same fashion, depending on the type of atoms.
The third term describes nearest neighbor hopping terms in the $[001]$ and equivalent crystal directions, the sum runs over all nearest neighbor pairs with $\br \in A$ and $\br' \in C$.
Depending on the atoms at sites $\br$ and $\br'$ the value of $V_{l, l', m, \br, \br'}$ is set to $V_{l, l', m}^{\text{SnTe}}$ if one of the sites is \ce{Sn} or $V_{l, l', m}^{\text{PbTe}}$ if one of the sites is \ce{Pb}.

All of the onsite energies and hopping terms are spin-independent, SOC only enters through the onsite SOC terms.
We summarize the parameter values used for numerical results in Table~\ref{lcao_table_18}.

Because of the identical outer shell electronic structure of \ce{Sn} and \ce{Pb}, the alloy composition $x$ does not affect the doping level, therefore, we set the Fermi level $E_F$ to ensure half filling for all compositions, see Fig.~\ref{fig:surface_and_mirror_chern} on the main text.

\begingroup
\squeezetable
\begin{table}
\begin{tabularx}{40mm}{X|SS}
    \toprule
     & \ce{SnTe} & \ce{PbTe} \\
    \midrule
$E_{sc}$  &  -6.578  &  -7.612 \\
$E_{sa}$  &  -12.067  &  -11.002 \\
$E_{pc}$  &  1.659  &  3.195 \\
$E_{pa}$  &  -0.167  &  -0.237 \\
$E_{dc}$  &  8.38  &  7.73 \\
$E_{da}$  &  7.73  &  7.73 \\
$\lambda_{pc}$  &  0.592  &  1.500 \\
$\lambda_{pa}$  &  0.564  &  0.428 \\
$V_{ss\sigma}$  &  -0.510  &  -0.474 \\
$V_{sp\sigma}$  &  -0.949  &  -0.705 \\
$V_{ps\sigma}$  &  0.198  &  -0.633 \\
$V_{pp\sigma}$  &  2.218  &  2.066 \\
$V_{pp\pi}$  &  -0.446  &  -0.430 \\
$V_{pd\sigma}$  &  -1.11  &  -1.29 \\
$V_{pd\pi}$  &  0.624  &  0.835 \\
$V_{dp\sigma}$  &  -1.67  &  -1.59 \\
$V_{dp\pi}$  &  0.766  &  0.531 \\
$V_{dd\sigma}$  &  -1.72  &  -1.35 \\
$V_{dd\delta}$  &  0.618  &  0.668 \\
\bottomrule
\end{tabularx}
\caption{Tight-binding parameters in electronvolts for \ce{SnTe} and \ce{PbTe} from \citet{Lent1986}. Note that we use opposite sign convention for $V_{sp\sigma}$ and $V_{ps\sigma}$. All other parameters not listed here vanish.}
\label{lcao_table_18}
\end{table}
\endgroup

\begin{figure}
    \includegraphics[width=0.75\columnwidth]{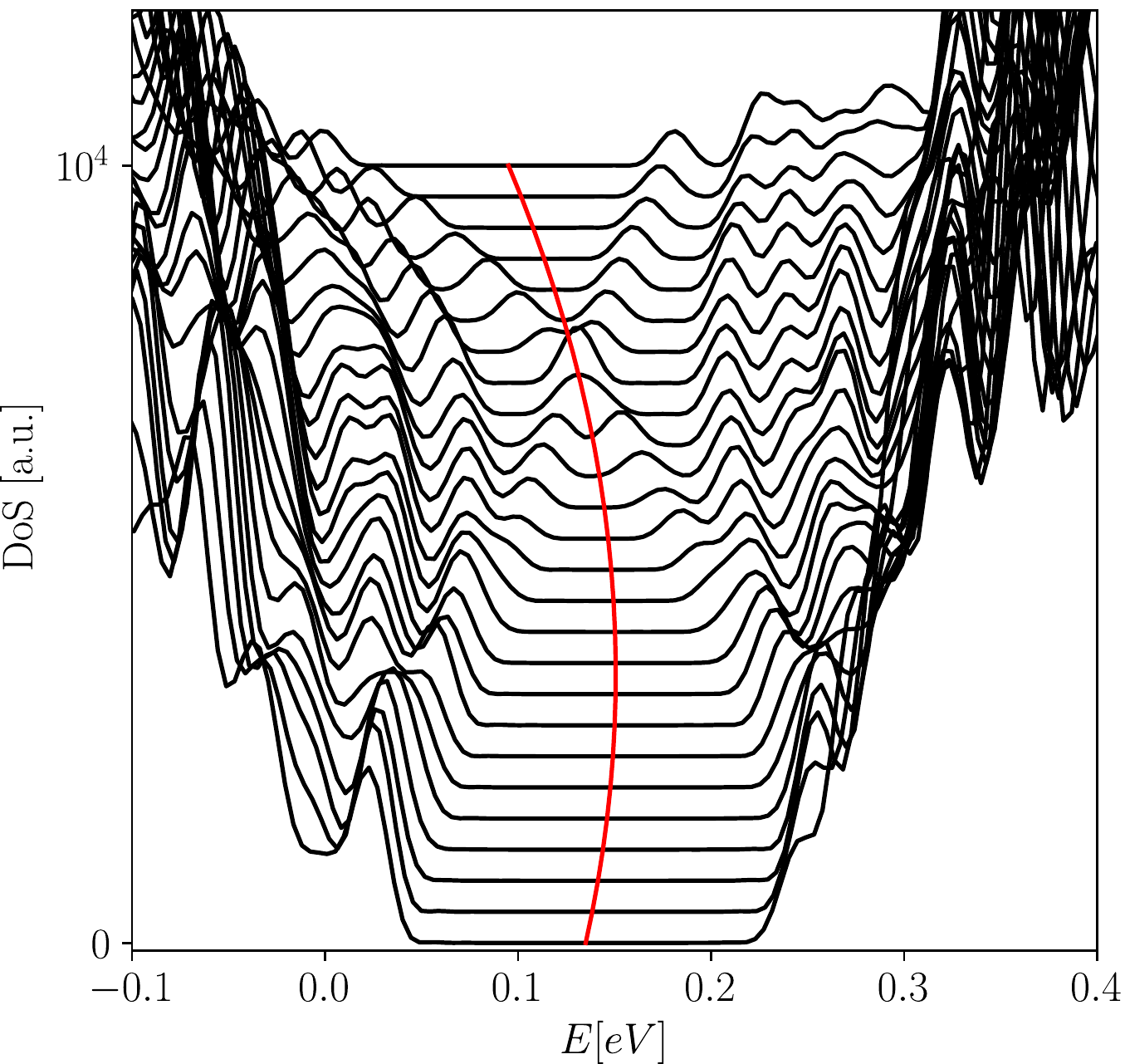}
    \caption{Density of states as function of $x$ near the bulk gap in the 18-band model of \alloy{}. The red line shows the placement of the Fermi level.}
    \label{fig:18band_DoS}
\end{figure}

\section{6-orbital tight-binding model}
\label{app:6orb}

\co{specify 6-orbital model}
We adopt the 6-orbital model of \ce{SnTe} and \ce{PbTe} originally described in~\citet{Mitchell1966}, and used in Refs.~\onlinecite{Hsieh2012, Fulga2016, Schindler2018, Sessi2016}.
Each site hosts spinful $p$-orbitals, 6 degrees of freedom in total with a vector of annihilation operators $\bc$. This Hamiltonian is formally identical to \eqref{eq:18band} but only includes $p$-orbitals and $pp\sigma$ hopping, while the hopping range is extended to second neighbors. The tight-binding Hamiltonian reads:
\begin{multline}
H=  \sum_{\br} m_{\br} \bc^{\dag}_{\br} \cdot \bc_{\br} + \sum_{\br} \lambda_{\br} \bc^{\dag}_{\br} \left(\bm{L} \cdot \bm{S}\right) \bc_{\br} \\
+\sum_{\left\langle\langle\br, \br'\right\rangle\rangle} t_{\br, \br'}  \bc^{\dag}_{\br'} \left[ \mathbbm{1} - \left(\hat{\bm{d}}_{\br,\br'}\cdot \bm{L} \right)^2\right] \bc_{\br}.
\end{multline}

The first term is the onsite energy (also termed ``mass term''), and it is the main source of disorder in our simulation.
$m_{\br}$ takes the value of $m_{\ce{Te}}$ on the $A$ sublattice, while for the $B$ sublattice a value is chosen between $m_{\ce{Sn}}$ and $m_{\ce{X}}$ with probability $1-x$ and $x$.
The second term is the ${\bm{L} \cdot \bm{S}}$ spin-orbit coupling, its value depends on the sublattice only (identical for \ce{Sn} and \ce{X}).
The third term is a $pp\sigma$ type of hopping \cite{Slater1954} that only connects $p$-orbitals oriented along the direction of the bond ${\hat{\bm{d}}_{\br,\br'}}$.

We include first neighbor $[001]$ and second neighbor $[110]$ hoppings, with amplitudes that depend on the sublattices.
We summarize the parameter values used for numerical results in Table~\ref{lcao_table_6}.

Because of the identical outer shell electronic structure of \ce{Sn} and \ce{Pb}, the alloy composition $x$ does not affect the doping level, therefore, we set the Fermi level $E_F$ to ensure half filling for all compositions.

\begingroup
\squeezetable
\begin{table}
\vspace{10pt}
\begin{tabularx}{25mm}{X|S}
    \toprule
     & \ce{SnTe} \\
    \midrule
$m_{\ce{Te}}$  &  -1.65 \\
$m_{\ce{Sn}}$  &  1.65 \\
$t_{aa}$  & -0.5 \\
$t_{ac} = t_{ca}$  &  0.9  \\
$t_{cc}$  &  0.5  \\
$\lambda_{a}$  &  -0.3 \\
$\lambda_{c}$  &  -0.3 \\
\bottomrule
\end{tabularx}
\caption{Tight-binding parameters in electronvolts for \ce{SnTe} used in the 6-orbital model. We use the same Hamiltonian as Ref.~\citet{Sessi2016}, but the numerical values of the parameters differ due to different normalization and sign conventions.}
\label{lcao_table_6}
\end{table}
\endgroup

\section{Evaluation and uncertainty of the finite-size scaling parameters}
\label{app:ffs}

\begin{figure}[ht]
    \includegraphics[width=0.95\columnwidth]{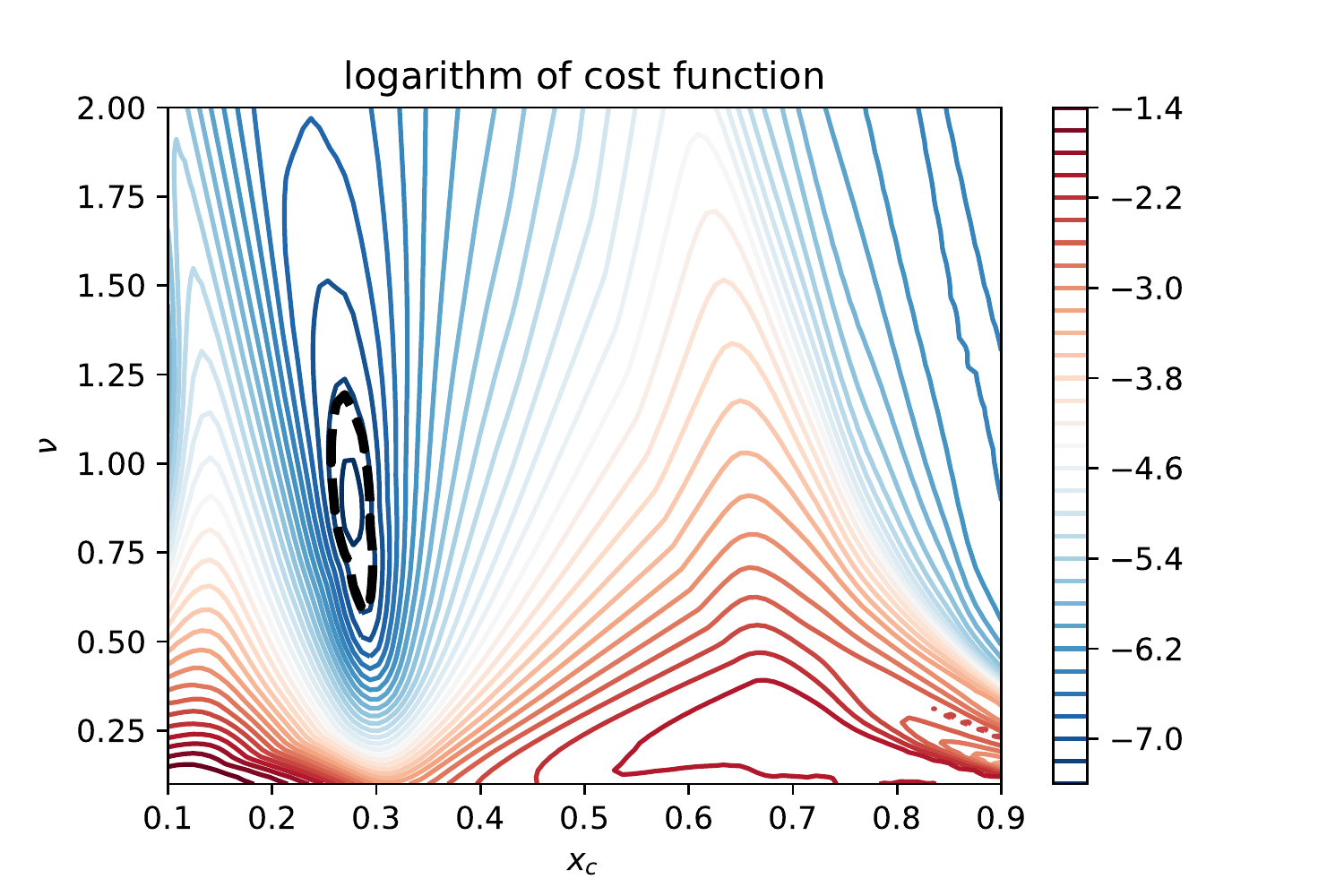}
    \caption{
    \label{log_cost_function}
    Logarithm of the cost function $\Phi(x_c, \nu)$ as a function of the parameters $(x_c, \nu)$.
    The region $D$ allowed by uncertainties around the minimum is delimited by a thick black dashed line.
    }
\end{figure}

We start from the data $C_M = f_i(x)$ of the the mirror Chern number as a function of the concentration of \alloy{} where $i=1,\dots,N$ correspond to different system sizes $L_i$. 
This data is obtained by averaging multiple disorder realizations.
We wish to perform a finite-size scaling collapse of the data.
The changes of variables $\tilde{x}_i = (x - x_c) L_i^{1/\nu}$ define new functions $\tilde{x} \mapsto \tilde{f}_i(\tilde{x})$, where both the critical concentration $x_c$ and the correlation length exponent $\nu$ are parameters to estimate so that the curves for different $L$ collapse onto a universal master curve.
Hence, we seek to minimize the distance between $\tilde{f}_i(\tilde{x})$ over their common domain of definition.
To do so, we interpolate the discrete data to perform the change of variable, evaluate $\tilde{f}_i(\tilde{x})$ over a fixed interval, and compute the variance of the data. 
This defines a cost function $\Phi(x_c, \nu)$ that is minimal at optimal parameters $(x_c^*, \nu^*) \simeq (\num{0.28}, \num{0.9})$.
The cost function is plotted as a function of the parameters $(x_c, \nu)$ in figure \ref{log_cost_function}.
To evaluate the uncertainty on the optimal parameters, we evaluate the cost function $\Phi(x_c^*, \nu^*)$ at the optimal parameters $(x_c^*, \nu^*)$ for different disorder realizations [that give different values of the initial data $C_{\text{M}, L}(x)$]. 
The standard deviation gives an estimation of the uncertainty $u(\Phi)$ on the cost function. 
The domain $D = \left\{  (x_c, \nu) \mid \Phi(x_c, \nu) \leq \Phi(x_c^*, \nu^*) + u(\Phi) \right\}$ where the cost function is closer to its minimum than the uncertainty $u(\Phi)$ is approximately an ellipse (see figure \ref{log_cost_function}) from which the uncertainties $u(x_c) \simeq \num{0.03}$ and $u(\nu) \simeq \num{0.6}$ can be directly estimated.

\end{document}